\documentclass[graybox]{svmult}

\usepackage{type1cm}        
\usepackage{makeidx}         
\usepackage{graphicx}        
\usepackage{multicol}        
\usepackage[bottom]{footmisc}

\usepackage{newtxtext}       %
\usepackage{newtxmath}       

\def\snn{\mbox{$\sqrt{s_{_{\rm NN}}}$}}

\newcommand{\bra}[1]{\langle #1|}
\newcommand{\ket}[1]{|#1\rangle}
\newcommand{\braket}[2]{\langle #1|#2\rangle}

\newcommand{\di}{{\rm d}}
\newcommand{\ii}{i}

\def\wJ{{\widehat J}}

\def\wP{{\widehat P}}
\def\wA{{\widehat A}}
\def\wB{{\widehat B}}
\def\wF{{\widehat F}}

\def\wLa{{\widehat \Lambda}}

\def\wrho{{\widehat{\rho}}}

\newcommand{\tr}{{\rm tr}}  
  
\newcommand{\e}{{\rm e}}

\newcommand{\p}{{\rm p}}

\newcommand{\be}{\begin{equation}}
\newcommand{\ee}{\end{equation}}                                                                               
\newcommand{\bea}{\begin{eqnarray}}
\newcommand{\eea}{\end{eqnarray}}

\makeindex             

\begin{document}

\title*{Strongly Interacting Matter under Rotation}
\author{Gaoqing Cao and Iurii Karpenko}
\institute{Gaoqing Cao \at Sun Yat-Sen University, Guangzhou 510275, China, \email{caogaoqing@mail.sysu.edu.cn}.
\and Iurii Karpenko \at Czech Technical University in Prague, B\v rehov\'a 7, 11519 Prague 1, Czech Republic, \email{yu.karpenko@gmail.com}.}
%
%
\maketitle
\section{Connecting Theory to Heavy Ion Experiment}\label{ThtoHIC}

\subsection{Feed-down effect of hyperon decays on the $\Lambda$ polarization}\label{Feeddown}

Only a fraction of all $\Lambda$ and $\bar{\Lambda}$ hyperons detected in heavy ion collisions are produced directly at
the hadronization stage thus are called {\em primary}. Indeed, a large fraction thereof stems from decays 
of heavier hyperons and one should account for the feed-down from higher-lying resonances when trying
to extract information about the vorticity from measurements of $\Lambda$ polarizations.
Particularly, the most important feed-down channels involve the strong decay $\Sigma^*\rightarrow
\Lambda + \pi$, the electromagnetic (EM) decay $\Sigma^0\rightarrow \Lambda +\gamma$, and the weak decay $\Xi
\rightarrow \Lambda + \pi$ \cite{Becattini:2016gvu}. Of course, there are also many heavier resonances which decay to either $\Sigma^0$ or $\Lambda$. As a matter of fact, in the heavy ion collisions of RHIC at energy $\sqrt{s_{\rm NN}}=200~{\rm GeV}$, the primary $\Lambda$ hyperons were predicted to contribute only a quarter to all that measured \cite{Becattini:2016gvu}. Therefore, the non-primary $\Lambda$ contributions from heavier hyperon decays dominate the final yield and may alter the polarization features of primary $\Lambda$. When polarized particles decay, their daughters are themselves polarized because of angular momentum conservation. In general, the fractions of polarization, which are inherited by the daughters or transferred from the mother to the daughters, depend on the momenta of the daughters in the rest frame of the mother.

Even though the theoretical predictions~\cite{Becattini:2016gvu} and experimental measurements are consistent with each other on the global polarization of $\Lambda$, that is, along the direction of the total angular momentum of fireball; they contradict with each other for the sign of either the transverse  local polarizations (TLPs) [with the global one excluded] or longitudinal local polarization (LLP), see the theoretical calculations with primary $\Lambda$~\cite{Karpenko:2016jyx,Becattini:2017gcx,Xia:2018tes,Florkowski:2019voj} and recent experimental measurements at STAR~\cite{Niida:2018hfw,Adam:2019srw}. Before any further great efforts are devoted to solving the "sign puzzles", one has to firstly check the simplest possibility: the feed-down effect from higher-lying hyperon decays on the $\Lambda$ polarization.
In this section, we're going to explore the feed-down effects on $\Lambda$ polarizations, especially the local ones. The theoretical derivations are mainly based on \cite{Becattini:2016gvu,Becattini:2019ntv,Xia:2019fjf} and the relevant experimental measurements were reported in \cite{Niida:2018hfw,Adam:2019srw}. All the secondary contributions to $\Lambda$ are two-body decays, thus we can generally denote the decay as $H\rightarrow \Lambda+X$, where $H$ and $X$ refer to the mother particle (heavier hyperon) and byproduct daughter particle (usually pions $\pi$ or photon $\gamma$), respectively. 

In the following, $H,\Lambda$ and $X$ will be simply called {\it Mother, Daughter} and {\it Byproduct}, so that the derivations and discussions can be generally applied to any other two-body fermion decays. Local thermodynamic equilibrium will be assumed for both the kinematics and spin dynamics of the system, and the small scattering interactions between hadrons will be neglected after the hadronization stage. Generally, three reference frames are involved in the study: the QGP frame (QGPF) which is the center of mass of the colliding nuclei in a collider experiment and the laboratorial observations base on, the Mother's rest frame (MRF) and the Daughter's rest frame (DRF). We assign different notations for the physical quantities in these frames: regular in the QGPF, subscript "$*$" in the MRF and subscript "$o$" in the DRF.

This section is arranged as follows. As a first trial, in Sec.\ref{Meanpolarization}, we derive the proportional coefficients for global polarization transfers from Mothers to Daughters by following the simple momentum-integrated formalism. The following sections mainly focus on the complicated local polarization transfers: In Sec.\ref{SDMM}, spin density matrix and derivation of momentum-dependent polarization are presented for the Mothers. Based on that, in Sec.\ref{Localpolarization}, we further use a formalism of reduced spin density matrix to calculate the local polarization of the Daughter in the decays, which is then weighted over the momentum distributions of the Mothers in Sec.\ref{Momentumaverage}. Finally, we compare our numerical calculations with the experimental measurements in Sec.\ref{puzzles}.

\subsubsection{Global polarization transfer to the Daughter} \label{Meanpolarization}

As long as one is interested in the momentum-integrated mean spin vector (MSV) of the Daughter in {\em its} own rest frame, we will show that a simple linear rule applies with respect to that of the Mother, that is,
\be\label{linear}
{\bf S}_{\Lambda o}= C_S~{\bf S}_{H*},
\ee
where $C_S$ is a coefficient which may or may not depend on the dynamical matrix elements. The proportionality between these two MSVs should be expected as, once the momentum integrations are carried out, the only special direction for the MSV of the Daughter is parallel to that of the MSV of the Mother. In many two-body decays, the conservation laws constrain the final state to such an extent that the coefficient $C_S$ is {\em independent} of the dynamical matrix elements. This happens, e.g., in the strong decay $\Sigma^*(1385) \to \Lambda \pi$ and the EM decay $\Sigma^0 \to \Lambda \gamma$, but not in the weak decay $\Xi \to \Lambda \pi$. Thus, this section will be devoted to determining the exact expressions of the coefficient $C_S$ for both strong and EM decays. For the exploration of global polarization transfer, the summations over all angular momentum components of the Daughter and Byproduct, $\lambda_\Lambda,\lambda'_\Lambda$ and $\lambda_X$, should be understood for both the MSVs and the normalization factors. For brevity, we will suppress the summation symbol "$\sum$ " over these indices in this section.

We will work out the exact relativistic results. In the relativistic framework, the use of the 
helicity basis is very convenient for complete descriptions of the helicity and alternative 
spin formalisms, we refer the readers to \cite{Moussa,Weinberg,Tung,Chung}. For the Mother, 
with spin $j$ and the $z$ component $m$ in its rest frame\footnote{In the rest frame, helicity coincides with the eigenvalue of the spin 
	operator $\widehat{S}$, conventionally $\widehat{S}_3$, see the textbooks.}, decaying into two particles 
$\Lambda$ and $X$, the final state $\ket{\psi}$ can be written as a superposition of states with 
definite momenta and helicities:
\begin{equation}
\ket{\psi} \propto \int \di \Omega_* \; D^j(\varphi_*,\theta_*,0)^{m*}_{\lambda}
\ket{{\bf p}_*,\lambda_\Lambda,\lambda_X} T^j(\lambda_\Lambda,\lambda_X).
\end{equation}  
Here, $\lambda = \lambda_\Lambda - \lambda_X$ with $\lambda_\Lambda$ and $\lambda_X$ the helicities of the daughters in the MRF, ${\bf p}_*$ is the three momentum of the Daughter $\Lambda$ with $\theta_*$ and $\varphi_*$ its spherical coordinates and
$\di \Omega_* = \sin\theta_* \di \theta_* \di \varphi_*$ the corresponding infinitesimal solid angle, $D^j$ is 
the Wigner rotation matrix in the representation of spin $j$, and $T^j(\lambda_\Lambda,\lambda_X)$ are the 
reduced dynamical amplitudes depending only on the final helicities.

Then, the relativistic MSV of the Daughter is given by:
$$
S^\mu_{\Lambda*} = \bra{\psi} \widehat{S}_{\Lambda*}^\mu \ket{\psi}
$$
with $\braket{\psi}{\psi}=1$, hence we have explicitly
\bea\label{spina1}
S^\mu_{\Lambda *}\! &=&\!{\int\! \di \Omega_* 
	D^j(\varphi_*,\theta_*,0)^{m*}_{\lambda}D^j(\varphi_*,\theta_*,0)^{m}_{\lambda'} \bra{\lambda'_\Lambda} \widehat{S}_{\Lambda *}^\mu \ket{\lambda_\Lambda} T^j(\lambda_\Lambda,\lambda_X)
	T^j(\lambda'_\Lambda,\lambda_X)^* \over \int \di \Omega_* \; 
	| D^j(\varphi_*,\theta_*,0)^{m*}_{\lambda} |^2 | T^j(\lambda_\Lambda,\lambda_X) |^2} \nonumber \\ 
\!&=&\!{\int\! \di \Omega_* 
	D^j(\varphi_*,\theta_*,0)^{m*}_{\lambda}D^j(\varphi_*,\theta_*,0)^{m}_{\lambda'} \bra{\lambda'_\Lambda} \widehat{S}_{\Lambda *}^\mu \ket{\lambda_\Lambda} T^j(\lambda_\Lambda,\lambda_X)
	T^j(\lambda'_\Lambda,\lambda_X)^* \over  \frac{4\pi}{2j+1}| T^j(\lambda_\Lambda,\lambda_X) |^2}.
\eea
According to our conventions, we emphasize that the numerator should sum over $\lambda_\Lambda,\lambda'_\Lambda$ and $\lambda_X$  and the denominator over $\lambda_\Lambda$ and $\lambda_X$, separately. To derive this expression, we have used the known results for the integrals of the Wigner $D$ matrices and the fact that the operator $\widehat{S}_{\Lambda*}$ does not change the momentum eigenvalue as well as the helicity of the Byproduct $X$. 

Now, the most important term in \eqref{spina1} is the transition amplitude of the spin operator: $\bra{\lambda'_\Lambda} \widehat{S}_{\Lambda *}^\mu \ket{\lambda_\Lambda}$. To evaluate it, we decompose the spin operator as the following:
$$\widehat{S}_{\Lambda *} = \sum_i \widehat{S}_{\Lambda *}^i n_i(p_*)$$
with $n_i(p_*)$ the three spacelike unit vectors orthogonal to the four-momentum $p_*$. Their expression can be obtained by applying the so-called 
{\em standard Lorentz transformation} $[p_*]$, which turns the unit time vector $\hat t$ into the direction of 
the four-momentum $p_*$ \cite{Moussa}, to the three spatial axis vectors ${\bf e}_i$, namely,
$$
n_i(p_*) = [p_*]({\bf e}_i);
$$  
so that
\be\label{spina2}
\widehat{S}_{\Lambda *} = [p_*]\left(\sum_i \widehat{S}_{\Lambda *}^i {\bf e}_i\right)
\ee
by taking advantage of the linearity of $[p_*]$. It is more convenient to rewrite the sum in the argument
of $[p_*]$ along the spherical vector basis:
\begin{eqnarray*}
	{\bf e}_{\pm}= \mp\frac{1}{\sqrt{2}}({\bf e}_1 \pm \ii {\bf e}_2), \	{\bf e}_0 = {\bf e}_3,
\end{eqnarray*}
upon which the $D^j$ matrix elements are defined. We have
\be\label{spina3}
\sum_i \widehat{S}_{\Lambda *}^i {\bf e}_i = - \frac{1}{\sqrt{2}} \widehat{S}_{\Lambda *}^- {\bf e}_+ + 
\frac{1}{\sqrt{2}} \widehat{S}_{\Lambda *}^+ {\bf e}_- + \widehat{S}_{\Lambda *}^0 {\bf e}_0=\sum_{n=-1}^1 a_n
\widehat{S}_{\Lambda*}^{-n}{\bf e}_n,  
\ee  
where $\widehat{S}_{\Lambda *}^\pm = \widehat{S}_{\Lambda *}^1 \pm i \widehat{S}_{\Lambda *}^2$ are the familiar spin ladder operators and $a_n = -n/\sqrt{2} + \delta_{n,0}$. The actions of these new operators onto the helicity ket $\ket{\lambda_\Lambda}$ are precisely the well-known ones onto the eigenstate of the $z$ component of angular momentum operator with eigenvalue $\lambda_\Lambda$, e.g.,
$$
\bra{\lambda_\Lambda'} \widehat{S}_{\Lambda *}^0 \ket{\lambda_\Lambda} = \lambda_\Lambda \delta_{\lambda_\Lambda,\lambda_\Lambda'}.
$$
Then, we can utilize (\ref{spina2}) and (\ref{spina3}) to rewrite the transition amplitude in a more explicit form as
\be\label{spina4}
\bra{\lambda'_\Lambda} \widehat{S}_{\Lambda *} \ket{\lambda_\Lambda} = \sum_{n=-1}^1 a_n
\bra{\lambda'_\Lambda} \widehat{S}_{\Lambda*}^{-n} \ket{\lambda_\Lambda} [p_*]({\bf e}_n).
\ee 

In order to work out (\ref{spina4}), we need to find an explicit expression for the standard transformation $[p_*]$.
In principle, it can be chosen freely but our choice treats $\lambda_\Lambda$ the Daughter's
helicity \cite{Weinberg,Tung}, then the expression is
\be\label{SLT}
[p_*] = {\sf R}_z(\varphi_*) {\sf R}_y(\theta_*) {\sf L}_z(\xi).
\ee
This is just a Lorentz boost along the $z$ axis with a hyperbolic 
angle $\xi$ such that $\sinh \xi = \|{\bf p}_*\|/m_\Lambda$, followed by a rotation around the $y$ axis with an angle $\theta_*$ 
and another one around the $z$ axis with an angle $\varphi_*$. Then,
\begin{eqnarray*}
	[p_*]({\bf e}_\pm) = {\sf R}_z(\varphi_*) {\sf R}_y(\theta_*)({\bf e}_\pm) = \sum_{l=-1}^1 D^1(\varphi_*,\theta_*,0)^l_{\pm 1} {\bf e}_l
\end{eqnarray*}
as ${\bf e}_\pm$ are Lorentz invariant under the boost along the $z$ axis. Conversely, ${\bf e}_0$ is not invariant under 
the Lorentz boost and transforms as
\begin{eqnarray*}
	[p_*]({\bf e}_0) &=& \cosh \xi {\sf R}_z(\varphi_*) {\sf R}_y(\theta_*)({\bf e}_0) + \sinh \xi {\sf R}_z(\varphi_*) 
	{\sf R}_y(\theta_*)(\hat t) \nonumber \\
	&=& \sum_{l=-1}^1 \frac{\varepsilon_{\Lambda*}}{m_\Lambda} D^1(\varphi_*,\theta_*,0)^l_0 {\bf e}_l + \frac{{\rm p}_*}{m_\Lambda} \hat t, 
\end{eqnarray*}
where ${\rm p}_*= \|{\bf p}_*\|$ and the energy $\varepsilon_{\Lambda*} = \sqrt{{\rm p}_*^2+m_\Lambda^2}$. By substituting these transformations into (\ref{spina4}), we eventually get the most explicit form
\bea\label{spina5}
\bra{\lambda'_\Lambda} \widehat{S}_{\Lambda *} \ket{\lambda_\Lambda} = \sum_{l,n} b_n
D^1(\varphi_*,\theta_*,0)^l_n \bra{\lambda'_\Lambda} \widehat{S}_{\Lambda*}^{-n} \ket{\lambda_\Lambda} {\bf e}_l + \lambda_\Lambda \delta_{\lambda_\Lambda,\lambda'_\Lambda} \frac{{\rm p}_*}{m_\Lambda} \hat t,
\eea 
where $b_n = -n/\sqrt{2} + \gamma_{\Lambda*} \delta_{n,0}$ with $\gamma_{\Lambda*}=\varepsilon_{\Lambda*}/m_\Lambda$ the Lorentz factor of
the Daughter.

We can now write down the fully expanded expression of the MSV $S_{\Lambda*}$
following (\ref{spina1}). The time component is especially simple: By using (\ref{spina5}), 
one has
\bea\label{timecomp}
S^0_{\Lambda *} =\frac{{\rm p}_*}{m_\Lambda} {\lambda_\Lambda \int \di \Omega_* \; 
	|D^j(\varphi_*,\theta_*,0)^{m*}_{\lambda}|^2 |T^j(\lambda_\Lambda,\lambda_X)|^2
	\over \frac{4\pi}{2j+1} |T^j(\lambda_\Lambda,\lambda_X) |^2},  
\eea  
which then reduces to
\be\label{timecomp2}
S^0_{\Lambda *} = \frac{{\rm p}_*}{m_\Lambda} \frac{
	\lambda_\Lambda |T^j(\lambda_\Lambda,\lambda_X) |^2}
{|T^j(\lambda_\Lambda,\lambda_X) |^2} 
\ee  
after carrying out the integral over $\Omega_*$ in the numerator. 
Similarly, the spatial components read:
\bea\label{spacecomp}
{\bf S}_{\Lambda *}&=&{T^j(\lambda_\Lambda,\lambda_X) 
	T^j(\lambda'_\Lambda,\lambda_X)^*\over \frac{4\pi}{2j+1}|T^j(\lambda_\Lambda,\lambda_X) |^2}\sum_{n,l} \bra{\lambda'_\Lambda} \widehat{S}_{\Lambda*}^{-n} \ket{\lambda_\Lambda}   \nonumber \\
&& \times b_n\int \di \Omega_* \; D^j(\varphi_*,\theta_*,0)^{m*}_{\lambda} D^j(\varphi_*,\theta_*,0)^{m}_{\lambda'}   
D^1(\varphi_*,\theta_*,0)^l_{n} {\bf e}_l. 
\eea
Since the analytic results for the angular integrals in (\ref{spacecomp}) are well known, the expression  can be greatly simplified in terms of Clebsch-Gordan coefficients:
\bea\label{spacecomp2}
{\bf S}_{\Lambda *}\!&=&\!{T^j(\lambda_\Lambda,\lambda_X) 
	T^j(\lambda'_\Lambda,\lambda_X)^* \sum_{n,l}\! b_n \bra{\lambda'_\Lambda} \widehat{S}_{\Lambda*}^{-n} \ket{\lambda_\Lambda}  \bra{jm} j1 \ket{ml} \bra{j \lambda} j1 \ket{\lambda'n}  {\bf e}_l \over T^j(\lambda_\Lambda,\lambda_X) |^2}\nonumber \\
\!&=&\!{T^j(\lambda_\Lambda,\lambda_X) 
	T^j(\lambda'_\Lambda,\lambda_X)^* \sum_{n}\!  b_n\bra{\lambda'_\Lambda} \widehat{S}_{\Lambda*}^{-n} \ket{\lambda_\Lambda}  \bra{jm} j1 \ket{m0} \bra{j \lambda} j1 \ket{\lambda'n}  {\bf e}_0\over T^j(\lambda_\Lambda,\lambda_X) |^2}.
\eea
Note that the only non-vanishing spatial component of the MSV ${\bf S}_{\Lambda *}$ is the one
along $z$ axis or proportional to ${\bf e}_0 = {\bf e}_3$. As the Mother is polarized along $z$ direction by construction, this is a consequence of rotational 
invariance. We'd like to mention that the integrands in both (\ref{timecomp}) and (\ref{spacecomp}), as functions of the angular variables $\theta_*$ and $\varphi_*$, are proportional to the MSV $S_{\Lambda *}({\bf p}_*)$ at some given momentum ${\bf p}_*$; see more details in the following sections. 

So far, what we have calculated is the MSV of the Daughter in the Mother's rest
frame. However, one is more interested in the MSV in the Daughter's rest 
frame. For a given momentum ${\bf p}_*$, this can be obtained by means of Lorentz boost:
$$
{\bf S}_{\Lambda o}({\bf p}_*) = {\bf S}_{\Lambda *}({\bf p}_*) - \frac{{\bf p}_*}{\varepsilon_{\Lambda*}(\varepsilon_{\Lambda*}+m_\Lambda)} {\bf S}_{\Lambda *}({\bf p}_*) \cdot {\bf p}_*.
$$
As $S_{\Lambda *}$ is a four-vector orthogonal to $p_*$ hence ${\bf S}_{\Lambda *}({\bf p}_*) \cdot {\bf p}_* = S^0_{\Lambda *}({\bf p}_*) \varepsilon_{\Lambda*}$ , we can evaluate the momentum-integrated MSV in DRF by following
\bea\label{ptransf}
{\bf S}_{\Lambda o} \equiv \langle {\bf S}_{\Lambda o}({\bf p}_*) \rangle =\langle {\bf S}_{\Lambda *}({\bf p}_*) \rangle  - \frac{\langle {\bf p}_* S^0_{\Lambda *}({\bf p}_*)\rangle}{\varepsilon_{\Lambda*}+m_\Lambda} 
= {\bf S}_{\Lambda *} - \frac{\langle {\bf p}_* S^0_{\Lambda *}({\bf p}_*)\rangle}{\varepsilon_{\Lambda*}+m_\Lambda}.
\eea
The first term on the right hand side is just the MSV shown in (\ref{spacecomp2}). While by adopting (\ref{timecomp}) and an alternative presentation of the momentum:
$$
{\bf p}_* = {\rm p}_* \sum_{l=-1}^1 D^1(\varphi_*,\theta_*,0)^l_0 {\bf e}_l,
$$
the second term can be evaluated according to
\bea\label{boostcomp}
\langle {\bf p}_* S^0_{\Lambda *}({\bf p}_*) \rangle &=&\frac{{\rm p}_*^2}{m_\Lambda} {\lambda_\Lambda 
	|T^j(\lambda_\Lambda,\lambda_X)|^2 \sum_{l=-1}^1 {\bf e}_l \int \di \Omega_* \; |D^j(\varphi_*,\theta_*,0)^{m*}_{\lambda}|^2 D^1(\varphi_*,\theta_*,0)^l_0 \over \frac{4\pi}{2j+1}| T^j(\lambda_\Lambda,\lambda_X) |^2}\ \nonumber \\
&=&\frac{{\rm p}_*^2}{m_\Lambda} { \lambda_\Lambda |T^j(\lambda_\Lambda,\lambda_X)|^2 \sum_{l=-1}^1 {\bf e}_l \bra{jm} j1 \ket{m l} \bra{j\lambda} j1 \ket{\lambda 0}\over | T^j(\lambda_\Lambda,\lambda_X) |^2}\nonumber\\
&=&\frac{{\rm p}_*^2}{m_\Lambda} { \lambda_\Lambda |T^j(\lambda_\Lambda,\lambda_X)|^2 \bra{jm} j1 \ket{m 0} \bra{j\lambda} j1 \ket{\lambda 0}\over | T^j(\lambda_\Lambda,\lambda_X) |^2} {\bf e}_0.
\eea
Then, by collecting both (\ref{spacecomp2}) and (\ref{boostcomp})  in (\ref{ptransf}), one finally gets
\bea\label{proper}
{\bf S}_{\Lambda o}={T^j(\lambda_\Lambda,\lambda_X) 
	T^j(\lambda'_\Lambda,\lambda_X)^* \sum_{n} c_n\bra{\lambda'_\Lambda} \widehat{S}_{\Lambda*}^{-n} \ket{\lambda_\Lambda}  \bra{jm} j1 \ket{m0} \bra{j \lambda} j1 \ket{\lambda'n}\over  | T^j(\lambda_\Lambda,\lambda_X) |^2} {\bf e}_0 
\eea
with the parameter
\be\label{cn}
c_n = - \frac{n}{\sqrt{2}} + \left( \gamma_{\Lambda*} - \frac{\beta_{\Lambda*}^2\gamma_{\Lambda*}^2}{\gamma_{\Lambda*}+1} \right) \delta_{n,0}
= - \frac{n}{\sqrt{2}} + \delta_{n,0}
\ee
the same as $a_n$. Note the disappearance of any dependence on the energy of the Daughter or the masses 
involved in the decay, once the MSV of the Daughter is boosted back to its rest frame; see also \eqref{linear} and \eqref{coeffic}. 

The MSV in (\ref{proper}) pertains to the Mother in state $\ket{jm}$, which
is a pure eigenstate of its spin operator $\widehat{S}_z$ in its rest frame. For a mixed state, the MSV should be weighted over the probabilities $P_m$ of different eigenstates\footnote{In the non-polarized case, $P_m$ is the same for any $m\in[-j,\dots,j]$}. Since it is known that
$$
\bra{jm} j1 \ket{m0} = \frac{m}{\sqrt{j(j+1)}}, 
$$
the weighted average turns out to be
\bea\label{polariz1}
{\bf S}_{\Lambda o} = \!\!\sum_{m=-j}^j \!mP_m  {\bf e}_0{T^j(\lambda_\Lambda,\lambda_X) T^j(\lambda'_\Lambda,\lambda_X)^* \sum_{n=-1}^1 c_n\bra{\lambda'_\Lambda} \widehat{S}_{\Lambda*}^{-n} \ket{\lambda_\Lambda} 
	\bra{j \lambda} j1 \ket{\lambda'n} \over \sqrt{j(j+1)}| T^j(\lambda_\Lambda,\lambda_X) |^2}.
\eea
It is easy to identify that $\sum_{m=-j}^j m P_m {\bf e}_0$ is just the MSV of the Mother  ${\bf S}_{H*}$, so
we finally verify that (\ref{linear}) holds, that is, the MSV of the Daughter in
DRF is proportional to that of the Mother in MRF. And the explicit form of the proportional coefficient is now clear:
\bea\label{coeffic}
C_S={T^j(\lambda_\Lambda,\lambda_X) T^j(\lambda'_\Lambda,\lambda_X)^* \sum_{n=-1}^1 c_n\bra{\lambda'_\Lambda} \widehat{S}_{\Lambda*}^{-n} \ket{\lambda_\Lambda} 
	\bra{j \lambda} j1 \ket{\lambda'n} \over \sqrt{j(j+1)}| T^j(\lambda_\Lambda,\lambda_X) |^2}.
\eea
According to the group theory, the Clebsch-Gordan coefficients involved in the estimation of (\ref{coeffic}) can be given directly as
\bea\label{cgc}
\bra{j \lambda} j1 \ket{\lambda 0}=\frac{\lambda}{\sqrt{j(j+1)}} , \
\bra{j \lambda} j1 \ket{(\lambda\mp 1) \pm 1}=\mp \sqrt{\frac{(j\mp\lambda +1)(j\pm\lambda)}{2j(j+1)}}. 
\eea
As has been mentioned before,
the somewhat surprising feature of (\ref{coeffic}) is that $C_S$ doesn't explicitly depend on the masses involved in the decay as $c_n$ is independent of them. There of course might be an implicit dependence on the masses through the dynamical amplitudes $T^j$, but this actually cancels out due to the normalization in several important instances.

If the decay is driven by parity-conserving interaction, such as the 
strong decay $\Sigma^* \rightarrow \Lambda \pi$ and EM decay $\Sigma^0 \rightarrow \Lambda \gamma$,
there is a known relation between the parity partners for the dynamical amplitudes \cite{Chung}:
\be\label{phase}
T^j(-\lambda_\Lambda,-\lambda_X) = \eta_H \eta_\Lambda \eta_X (-1)^{j-S_\Lambda-S_X} 
T^j(\lambda_\Lambda,\lambda_X).
\ee 
Here, $\eta_H,\eta_\Lambda$ and $\eta_X$ are the intrinsic parities of the Mother, Daughter and Byproduct, and $j,S_\Lambda$ and $S_X$ are their spins, respectively. Note that the helicity is constrained to $\lambda_X=\pm S_Xß$ in (\ref{phase}) if the Byproduct is massless \cite{Tung}. In all these cases, one has
\be\label{phase2} 
|T^j(-\lambda_\Lambda,-\lambda_X)|^2 = |T^j(\lambda_\Lambda,\lambda_X)|^2.
\ee
The (\ref{phase}) and (\ref{phase2}) have interesting consequences: First of all, 
it can be readily realized that the time component of the MSV (\ref{timecomp2}) vanishes. 
Secondly, if only one dynamical amplitude is independent in (\ref{coeffic}) because of the constraint from (\ref{phase}), the coefficient $C_S$ can be finally reduced to a constant that is determined 
only by the conservation laws. We will see below that this is precisely the case for the decays $\Sigma^* \rightarrow \Lambda \pi$ 
and $\Sigma^0 \rightarrow \Lambda \gamma$.\\

\paragraph{{\textbf{A. Strong decay $\Sigma^* \rightarrow \Lambda \pi$}}}

In this case, $\lambda_X = 0$, $\lambda =\lambda_\Lambda$, $j=3/2$ and $T^j(\lambda)$ is proportional to
$T^j(-\lambda)$ through a phase factor, which turns out to be $1$ according to (\ref{phase}). As 
$\lambda_\Lambda=\pm1/2$, there is only one independent reduced helicity amplitude thus the coefficient
$C_S$ simplifies to
\be\label{sigma1}
C_S =\sum_{n=-1}^1 \sum_{\lambda,\lambda'}\bra{\lambda'} \widehat{S}_{\Lambda*}^{-n} \ket{\lambda} 
\frac{c_n}{\sqrt{j(j+1)}} \frac{\bra{j \lambda} j1 \ket{\lambda'n}}{2S_\Lambda + 1}.
\ee
We now evaluate the three terms in the above summation over $n$ one by one. For $n=0$, one 
obtains
$$
\sum_{\lambda=\pm1/2}\frac{1}{2} \lambda^2 \frac{1}{j(j+1)} = \frac{1}{15},
$$
where the first equation in (\ref{cgc}) has been used. For $n=1$, the corresponding ladder operator in (\ref{sigma1}) is $\widehat{S}_{\Lambda *}^-$, which selects the term with $\lambda'=-1/2$ 
and $\lambda=1/2$ as the only non-vanishing contribution. Similarly, for $n=-1$, the corresponding ladder operator $\widehat{S}_{\Lambda *}^+$ in (\ref{sigma1})
selects the opposite combination: $\lambda'=1/2$ 
and $\lambda=-1/2$. According to the second equation in (\ref{cgc}), the corresponding Clebsch-Gordan coefficients are opposite to each other for $n=\pm1$. Then, by inserting (\ref{cn}), their contributions turn out to be the same, that is,
$$
\frac{1}{2} \sqrt{\frac{8}{15}} \frac{1}{\sqrt{2}} \frac{1}{\sqrt{j(j+1)}} = \frac{2}{15}.
$$
Therefore, the coefficient $C_S$ is just
\be\label{sigma2}
C_S = \frac{1}{15} + 2 \frac{2}{15} = \frac{1}{3},
\ee
which indicates that the MSV of the Daughter is along that of the Mother. 

\paragraph{{\textbf{B. Electromagnetic decay $\Sigma^0 \rightarrow \Lambda \gamma$}}}

This case is fully relativistic as the Byproduct is a photon, then the helicity basis 
is compelling with $\lambda_X=\pm 1$. Now $j=1/2$, (\ref{spina1}) indicates that
$$
|\lambda| = |\lambda_\Lambda - \lambda_X| = 1/2,
$$
thus only two choices are possible:
\begin{eqnarray*}
	& \lambda_X =  1 \implies \lambda_\Lambda = 1/2  \implies \lambda = -1/2&, \nonumber \\
	& \lambda_X = -1 \implies \lambda_\Lambda = -1/2  \implies \lambda = 1/2&,
\end{eqnarray*}
from which we can generally identify $\lambda_X = 2\lambda_\Lambda$ and $\lambda = -\lambda_\Lambda$ in (\ref{coeffic}). 
The same argument applies to $\lambda' = \lambda'_\Lambda -\lambda_X$, so we also have $\lambda_X
= 2 \lambda'_\Lambda$, whence $\lambda'_\Lambda = \lambda_\Lambda$ and $\lambda' = \lambda$. This in turn implies that only the term with $n=0$ contributes in (\ref{coeffic}), which then reads
\bea\label{sigmaz1}
&& C_S = {\lambda_\Lambda |T^j(\lambda_\Lambda,2\lambda_\Lambda)|^2 \bra{j -\lambda_\Lambda} j1 \ket{-\lambda_\Lambda 0} \over \sqrt{j(j+1)}| T^j(\lambda_\Lambda,2\lambda_\Lambda) |^2}.
\eea
Like the previous case, there is only one independent dynamical amplitude because of the constraint from (\ref{phase2}), so (\ref{sigmaz1}) becomes
\be\label{sigmaz2}
C_S = \sum_{\lambda_\Lambda=\pm1/2} \lambda_\Lambda \frac{(-\lambda_\Lambda)}{j(j+1)}\frac{1}{2 S_\Lambda +1}
\ee
by inserting the first equation in (\ref{cgc}). With the spins $j=S_\Lambda=1/2$, we eventually recover the known result~\cite{Cha,Armenteros}:
$$
C_S=-\frac{1}{3},
$$
which indicates that the MSV of the Daughter is along the opposite direction to that of the Mother.

\subsubsection{Spin density matrix for the Mother and its polarization}\label{SDMM}

In general, the Mother's eigenstates do not have definite spins in a local thermodynamic 
equilibrium (LTE) system with angular momentum-vorticity coupling, which means that the Mother's spin can be altered. To account for the spin transition, the spin density matrix (SDM) can be defined: For the Mother with four-momentum $p_H$ in QGPF, the form is given by
\be\label{spindens}
\Theta(p_H)_{\sigma \sigma'} = \frac{\tr (\wrho a^\dagger(p_H)_{\sigma'} a(p_H)_{\sigma})}
{\sum_\sigma \tr (\wrho a^\dagger(p_H)_{\sigma} a(p_H)_\sigma)},
\ee
where $a^\dagger(p_H)_{\sigma}$ and $a(p_H)_\sigma$ are creation and annihilation operators of the Mother in the spin state $\sigma$, respectively. As mentioned before, the meaning of $\sigma$ depends on the choice of the {\em standard Lorentz transformation} $[p_H]$ which transforms $\hat{t}$ to the direction of $p_H$~\cite{Moussa}. For convenience and consistency, we adopt the choice that $\sigma$ stands for the particle's helicity~\cite{Tung} in the following. Similar to that for the Daughter \eqref{SLT}, the transformation is explicitly given by:
\be\label{standard}
[p_H] = {\sf R}(\varphi,\theta,0) {\sf L}_z(\xi) ={\sf R}_z(\varphi) {\sf R}_y(\theta) {\sf L}_z(\xi),
\ee
where the functions have the same meanings as those in \eqref{SLT} but is for the Mother in  QGPF here.

Then, by operating the transformation over the space-like orthonormal vector basis ${\bf e}_i$ as $n_i(p_H) \equiv [p_H]({\bf e}_i)$~\cite{Becattini:2016gvu,Moussa}, one can readily determine the MSV of the Mother from the SDM (\ref{spindens}) as:
\begin{eqnarray}\label{meanspin}
S_H^\mu(p_H) = \sum_{i=1}^3 \tr \left[D^j({\sf J}^i)\Theta(p_H)\right]\,n_i(p_H)^\mu
= \sum_{i=1}^3 [p_H]^\mu_i \tr \left[D^j({\sf J}^i)\Theta(p_H)\right], 
\end{eqnarray}
where ${\sf J}^i$ are the angular momentum generators of the Mother and $D^j({\sf J}^i)$ their irreducible
representation matrices with total spin $S$. It should be stressed that, in spite of the appearance of the Lorentz transformation $[p_H]$, the MSV is independent of its particular choice as should be for any observables. Actually, the SDM \eqref{spindens} also depends on the convention of $[p_H]$ implicitly through the definition of the spin variable $\sigma$, which just compensates the explicit dependence. By adopting the covariant form of the irreducible representation matrix
\be\label{dspin}
D^j({\sf J}^\lambda) = -\frac{1}{2} \epsilon^{\lambda\mu\nu\rho} D^j(J_{\mu\nu})\hat t_\rho,
\ee
which indicates $D^j({\sf J}^0)=0$ for the unit time vector $\hat t=(1,0,0,0)$, the MSV \eqref{meanspin} can be conveniently rewritten with the full Lorentz covariant indices as:
\be\label{meanspincov}
S_H^\mu(p_H) = [p_H]^\mu_\nu \tr \left[D^j({\sf J}^\nu) \Theta(p_H)\right]. 
\ee

Now, the most important mission is to evaluate the SDM for a general spin $S$, which is not an easy task in quantum field theory (QFT): Even for the simplest non-trivial case with the density operator involving the angular 
momentum-vorticity coupling, an exact solution is unknown. However, it is possible to find
an explicit exact solution for a single spicy of relativistic quantum particles by neglecting quantum
statistic (or quantum field) effects. In this case, the general density 
operator $\wrho$ for a system in  equilibrium is given by
$$
\wrho = \frac{1}{Z} \exp \left[-b \cdot \wP + \frac{1}{2} \varpi : \wJ\right],
$$
where $b$ is a time-like constant four-vector, $\varpi$ an anti-symmetric constant tensor, and $\wP$ and $\wJ$ are the conserved total four-momentum and total angular momentum operators, respectively. As the scattering effects are neglected in our study, the system can be viewed as a set of non-interacting distinguishable particles. Then we can write
$$
\wP = \sum_{i} \wP_i, \qquad \qquad  \wJ = \sum_i \wJ_i,
$$
and consequently $\wrho = \otimes_i \wrho_i$ with the density operator for a single particle specy
$$
\wrho_i = 
\frac{1}{Z_i} \exp \left[-b \cdot \wP_i + \frac{1}{2} \varpi : \wJ_i\right].
$$

By following the Poincar\'e group algebra for the generators of translations $\wP_\mu$ and Lorentz transformations $\wJ_{\mu\nu}$~\cite{Weinberg,Tung}:
\be\label{commutators}
[\wP_\mu,\wP_\nu]=0, \qquad \qquad [\wP_\tau,\wJ_{\mu\nu}]=-i(\wP_\mu \eta_{\nu\tau}-\wP_\nu \eta_{\mu\tau}),
\ee
it can be shown that
\bea
\wF_{kj}&\equiv&\left[\left[(-b \cdot \wP_i),\left(\frac{1}{2} \varpi : \wJ_i\right)^{(k)}\right],(-b \cdot \wP_i)^{(j)}\right]\nonumber\\
&=&-(-i)^k\wP_i^\mu\underbrace{\left( 
	\varpi_{\mu\nu_1} \varpi^{\nu_1\nu_2} \ldots \varpi_{\nu_{k-1}\nu_k} \right)}_\text{k times}b^{\nu_k} \delta_{j0},
\eea
where $\left[\wA,\wB^{(k)}\right]=\left[\left[\cdots\left[\wA,\wB\right],\cdots\right],\wB\right]$ with $k$ times of nesting commutations and $\wF_{kj}$ commute with each other for any $k$ and $j$. In the most general case with arbitrary operators $\wA$ and $\wB$, an identity has been derived: 
$$e^{\wA+\wB}=e^\wA e^\wB e^{-{1\over2}[\wA,\wB]}e^{{1\over6}[\wA^{(2)},\wB]-{1\over3}[\wA,\wB^{(2)}]}\cdots,$$
 where the higher level commutation exponents in "$\cdots$" rely on the lower ones through some recursion relations~\cite{Lin:2002}. In our present case, all the non-vanishing commutation exponents must be functions of $\wP_\mu$ according to \eqref{commutators} and commute with each other. Thus, a general identity can be applied to the density operator and we have~\cite{Lin:2002}
\bea
\wrho_i=\frac{1}{Z_i} \exp\left\{\sum_{k=1}^\infty{(-1)^k\wF_{k0}\over (k+1)!}\right\}\exp [-b \cdot \wP_i] \exp \left[\frac{1}{2} \varpi : \wJ_i\right].
\eea
Then, it can be rewritten in a very simple factorized form as:
\be\label{densi}
\wrho_i = \frac{1}{Z_i} \exp [-\tilde b \cdot \wP_i] \exp \left[\frac{1}{2} \varpi : \wJ_i\right]
\ee
by defining a $\varpi$ dependent effective four vector
$$
\tilde b_\mu = \sum_{k=0}^\infty \frac{\ii^k}{(k+1)!} \underbrace{\left( 
	\varpi_{\mu\nu_1} \varpi^{\nu_1\nu_2} \ldots \varpi_{\nu_{k-1}\nu_k} \right)}_\text{k times} 
b^{\nu_k}. 
$$

As the non-commutative operators $\wP_i$ and $\wJ_i$ are completely separated from each other into two independent multiplying exponential functions in \eqref{densi}, the SDM for the Mother can be reduced to a simple form:
\be\label{spindens2}
\Theta(p_H)_{\sigma \sigma'} = \frac{\bra{p_H,\sigma} \wrho_H \ket{p_H,\sigma'}}
{\sum_\sigma \bra{p_H,\sigma} \wrho_H \ket{p_H,\sigma}}=\frac{\bra{p_H,\sigma} \exp \left[\frac{1}{2} \varpi : \wJ_H\right]\ket{p_H,\sigma'}}
{\sum_\sigma \bra{p_H,\sigma} \exp \left[\frac{1}{2} \varpi : \wJ_H\right] \ket{p_H,\sigma}}.
\ee
It is now completely determined by its single particle density operator $\wrho_H$ or more precisely the angular momentum dependent part.

To derive the explicit form for (\ref{spindens2}), we use a convenient analytic continuation technique: we first
derive $\Theta(p_H)$ for imaginary $\varpi$ and then continue the result back to real
value. In the former case, $\wLa \equiv \exp[\varpi:\wJ_H/2]$ is just a unitary 
representation of Lorentz transformation, then the well-known relations in group theory can be used to
obtain:
\be\label{spindens3}
\Theta(p_H)_{\sigma \sigma'} = \frac{\bra{p_H,\sigma} \wLa \ket{p_H,\sigma'}}
{\sum_\sigma \bra{p_H,\sigma} \wLa \ket{p_H,\sigma}} = 
\frac{ W(p_H)_{\sigma\sigma'}2 \varepsilon_H\delta^3({\bf p}_H - {\bf \Lambda}({\bf p}_H))}
{W(p_H)_{\sigma\sigma}2 \varepsilon_H\delta^3({\bf p}_H - {\bf \Lambda}({\bf p}_H))}.
\ee
Here, ${\bf \Lambda}({\bf p}_H)$ stands for the spatial part of the four-vector $\Lambda(p_H)$ and the covariant normalization scheme is used for the Mother eigenstates, that is,
$$
\braket{p_H,\sigma}{p_H', \sigma'} = 2 \varepsilon_H \delta^3({\bf p}_H-{\bf p}'_H)
\delta_{\sigma\sigma'}.
$$
In (\ref{spindens3}), the matrix $W(p_H)$ is the so-called Wigner rotation matrix:
$$
W(p_H) = D^j([{\Lambda}p_H]^{-1} {\Lambda} [p_H]),
$$
where $D^j$ is the $(2S+1)$-dimensional representation, the
so-called $(0,2S+1)$~\cite{Tung}, of the SO(1,3)-SL(2,C) matrices in the argument 
\cite{Becattini:2019ntv}.
Altogether, the SDM for the Mother is simply
\be\label{densiIO}
\Theta(p_H)_{\sigma \sigma'} = \frac{D^j([p_H]^{-1} {\Lambda} [p_H])_{\sigma\sigma'}}
{\tr \left[D^j({ \Lambda})\right]},
\ee
which seems appropriate to be analytically continued to real $\varpi$. 

However, it is not satisfactory yet as the analytic continuation of \eqref{densiIO} to real $\varpi$, that is,
\be\label{analcont}
D^j({\Lambda}) \to \exp\left[\frac{1}{2} \varpi : J_H\right],
\ee
does not give rise to a hermitian matrix for $\Theta(p_H)$ as it should. This problem 
can be fixed by taking into account the fact that $W(p_H)$ is the representation of a rotation
hence unitary. We thus replace $W(p_H)$ with $(W(p_H) + W(p_H)^{-1\dagger})/2$ in 
(\ref{spindens3}) and obtain, by using the transparency to the adjoint operation property of SL(2,C) representations,
$$
\Theta(p_H) = \frac{D^j([p_H]^{-1} { \Lambda} [p_H])+
	D^j([p_H]^{\dagger} { \Lambda}^{-1 \dagger}[p_H]^{-1\dagger})}
{\tr \left[D^j({ \Lambda})+D^j({ \Lambda})^{-1 \dagger}\right]}.
$$
As the analytic continuation of ${\Lambda}^{-1\dagger}$ related part reads
\be\label{analcont2}
D^j({\Lambda}^{-1\dagger}) \to \exp\left[\frac{1}{2} \varpi :  D^{j\dagger}(J)\right],
\ee
the final expression of the SDM in a rotational system is:
\be\label{spindensf}
\Theta(p_H) = \frac{\sum_{{\cal O}=1,\dagger}\left[D^j([p_H]^{-1} \exp[\varpi :  D^{j}(J)/2] [p_H])\right]^{\cal O}}
{\tr \left[\exp[\varpi :  D^{j}(J)/2] + \exp[\varpi :  D^{j\dagger}(J)/2]\right]},
\ee
which is manifestly hermitian. 

The expression can be further simplified: By taking the involved matrices as
SO(1,3) transformations and using known relations in group theory, we have
$$
[p_H]^{-1} \exp \left[ \frac{1}{2} \varpi : J\right] [p_H] = \exp\left[ 
\frac{1}{2} \varpi^{\mu\nu} [p_H]^{-1} J_{\mu\nu} [p_H] \right] =  
\exp\left[ \frac{1}{2} \varpi^{\alpha\beta}_{*}(p_H) J_{\alpha\beta}
\right],
$$
where the effective anti-symmetric tensor $\varpi_*$ is defined as:
\be\label{varpiboost}
\varpi^{\alpha\beta}_{*}(p_H)\equiv\varpi^{\mu\nu} [p_H]^{-1\alpha}_\mu [p_H]^{-1 \beta}_\nu.
\ee
Actually, $\varpi_*^{\alpha\beta}$ have physical meanings themselves, that is, the components of thermal vorticity tensor 
in the MRF. They are obtained  from the ones in QGPF
by taking the inverse transformation of $[p_H]$. Finally, 
(\ref{spindensf}) becomes
\be\label{spindensf2}
\Theta(p_H) = \frac{D^j(\exp[\varpi_*(p_H) :  D^{j}(J)/2])+ D^j(\exp[\varpi_*(p_H) : D^{j\dagger}(J)/2])}
{\tr (\exp[\varpi :  D^{j}(J)/2]) + \exp[\varpi : D^{j\dagger}(J)/2])}.
\ee

In many cases, such as in peripheral heavy ion collisions, the thermal vorticity $\varpi$ is usually $\ll 1$ due to the relatively large proper temperature~\cite{Becattini:2017gcx}, so the SDM can be
expanded in power series around $\varpi=0$. Take into account the traceless of the generators of Lorentz transformation, that is $\tr(J_H)=0$, we have:
$$
\Theta(p_H)^\sigma_{\sigma'} \simeq \frac{\delta^\sigma_{\sigma'}}{2j+1} + \frac{1}{4(2j+1)} 
\varpi_*^{\mu\nu}(p_H)\left(D^{j}(J_{\mu\nu})+D^{j\dagger}(J_{\mu\nu})\right)^\sigma_{\sigma'}
$$
to the order $o(\varpi)$. The representation $D^j(J_{\mu\nu})$ can be decomposed as the following:
\be\label{dspin2}
D^j(J_{\mu\nu}) = \epsilon_{\mu\nu\rho\tau} D^j({\sf J}^\rho)\hat t^\tau
+ D^j({\sf K}_\nu)\hat t_\mu - D^j({\sf K}_\mu)\hat t_\nu 
\ee
with $D^j({\sf J}^i)$ hermitian and $D^j({\sf K}^i)$ anti-hermitian matrices, respectively. Then, we find that the SDM is only rotation relevant:
\be\label{spindensexp}
\Theta(p_H)^\sigma_{\sigma'} \simeq \frac{\delta^\sigma_{\sigma'}}{2j+1} + \frac{1}{2(2j+1)} 
\varpi_*^{\mu\nu}(p_H) \epsilon_{\mu\nu\rho\tau} D^j({\sf J}^\rho)^\sigma_{\sigma'} 
\hat t^\tau.
\ee
Note that the number of the generators $D^{S}(J)$ is more than three for $S>1/2$, but only three is involved in \eqref{spindensexp} with the others functioning through higher order terms of $\varpi$.
By substituting \eqref{spindensexp} into \eqref{meanspincov}, only the second term of \eqref{spindensexp} contributes:
\begin{align}\label{vortspin}
S_H^\mu(p_H) &= [p_H]^\mu_\kappa \frac{1}{2(2j+1)} \varpi_*^{\alpha\beta} (p_H)
\epsilon_{\alpha\beta\rho\tau} \tr \left( D^j({\sf J}^\rho) D^j({\sf J}^\kappa) \right) 
\hat t^\tau \nonumber \\
& = -\frac{j(j+1)}{6} [p_H]^\mu_\rho \varpi_{*\alpha\beta} (p_H)
\epsilon^{\alpha\beta\rho\tau} \hat t_\tau = -\frac{j(j+1)}{6m_H} \epsilon^{\alpha\beta\mu\tau}\varpi_{\alpha\beta} 
 p_{H\tau},
\end{align}
where we have transformed back to the QGPF by inserting \eqref{varpiboost} in the last equality. As we will see in next section, the MSV can be boosted to the MRF to give the true spin observables ${\bf S}_{H*}$ and then the polarization of the Mother is defined as ${\bf P}_H={\bf S}_{H*}/j$.

\subsubsection{Local polarization transfer to the Daughter}\label{Localpolarization}

Now, with the Mother's local polarization determined in the previous section, it's the right time to study the polarization transfer to the Daughter from the feed-down effect of the Mother in two-body decays. Concretely, the most important mission is to derive the reduced spin density matrix for the Daughter in the MRF by tracing over the quantum states of the Byproduct, which thus indicates that this reduced SDM should be mixed rather than pure in general. In the MRF, the magnitude of the three-momentum of the Daughter is fixed due to energy-momentum conservation, that is,
\be\label{pmagn}
p_{*}=p_{\Lambda*}\equiv{1\over2{m_H}}\prod_{s,t=\pm}({m}_H +s\,m_{\Lambda}+t\,m_{X})^{1/2}.
\ee
As mentioned in Sec.~\ref{Meanpolarization}, as long as the decay hasn't been observed, contribution of the Mother state to the quantum superposition of the Daughter and Byproduct reads in the helicity basis as
\cite{Moussa,Tung,Chung}
\be\label{state}
\ket{p_* j m \lambda_\Lambda \lambda_X} \propto  T^j(\lambda_\Lambda,\lambda_X) \int \di \Omega_* \; 
D^j(\varphi_*,\theta_*,0)^{m \, *}_{\lambda} \ket{{\bf p}_* \lambda_\Lambda\lambda_X}.
\ee
Once a measurement is made for the momentum of either final particle hence ${\bf p}_*$ is fixed down, we can define the non-integrated form of the two-body spin density operator as\footnote{For brevity, the summation convention is assumed: if an angular momentum component 
	index (only for superscripts and subscripts) shows more than once in the formula, the index should summed over. For example, we should sum 
	over $m$ in the numerator of (\ref{rho1}) and over $m,\lambda_\Lambda$ and $\lambda_X$ in the denominator as $|D^j(\varphi,\theta,0)^{m}_{\lambda}|^2=D^j(\varphi,\theta,0)^{m \, *}_{\lambda_\Lambda-\lambda_X}D^j(\varphi,\theta,0)^{m}_{\lambda_\Lambda-\lambda_X}$.}
\be\label{rho1}
\wrho({\bf p}_*) = \frac{T^j(\lambda_\Lambda,\lambda_X) T^j(\lambda'_\Lambda,\lambda'_X)^* 
	D^j(\varphi_*,\theta_*,0)^{m \, *}_{\lambda} D^j(\varphi_*,\theta_*,0)^{m}_{\lambda'}
	\ket{{\bf p}_* \lambda_\Lambda\lambda_X} \bra{{\bf p}_* \lambda'_\Lambda \lambda'_X}}
{|T^j(\lambda_\Lambda,\lambda_X)|^2 |D^j(\varphi,\theta,0)^{m}_{\lambda}|^2
	\braket{{\bf p}_* \lambda_\Lambda \lambda_X}{{\bf p}_* \lambda_\Lambda \lambda_X}}
\ee
for a given state of the Mother with $z$ component of spin $m$. 

However, as we've illuminated in the previous section, the spin state of the Mother can be shifted according to \eqref{spindensf2} in a rotational system. In this case, the two-body density operator
should be a mixing of different spin states of the Mother:
\be\label{state2}
\sum_{m,n=-j}^j \Theta(p_H)^m_{n} \ket{p_* jm \lambda_\Lambda \lambda_X}\bra{p_* j n \lambda'_\Lambda \lambda'_X},
\ee
rather than the pure one with $\Theta(p_H)^m_{n}\rightarrow\delta^m_{n}$. To be consistent with the setup, the involved matrices $D^j(J)$ in \eqref{spindensf2} are now also defined in the MRF, which then allows us to apply the usual matrix algebra in later explicit evaluations. Following \eqref{state2}, a more general density operator for the daughters with fixed momentum ${\bf p}_*$ reads:
\bea\label{rho2}
\wrho({\bf p}_*) &\propto& T^j(\lambda_\Lambda,\lambda_X) T^j(\lambda'_\Lambda,\lambda'_X)^* 
D^j(\varphi_*,\theta_*,0)^{m \, *}_{\lambda} \Theta(p_H)^m_{n} D^j(\varphi_*,\theta_*,0)^{n}_{\lambda'}\nonumber\\
&&\ket{{\bf p}_* \lambda_\Lambda\lambda_X} \bra{{\bf p}_* \lambda'_\Lambda \lambda'_X}.
\eea
Then, the normalized two-body spin density matrix follows directly:
\be\label{spindensmat}
\Theta(\varphi_*,\theta_*)_{\lambda'_\Lambda \lambda'_X}^{\lambda_\Lambda \lambda_X} = 
\frac{T^j(\lambda_\Lambda,\lambda_X) T^j(\lambda'_\Lambda,\lambda'_X)^* D^j(\varphi_*,\theta_*,0)^{m \, *}_{\lambda} \Theta(p_H)^m_{n} D^j(\varphi_*,\theta_*,0)^{n}_{\lambda'}}
{|T^j(\lambda_\Lambda,\lambda_X)|^2 D^j(\varphi_*,\theta_*,0)^{m \, *}_{\lambda} 
	\Theta(p_H)^m_{n} D^j(\varphi_*,\theta_*,0)^{n}_{\lambda}}.
\ee

Combining \eqref{meanspincov} and \eqref{spindensmat}, the MSV of the Daughter can be obtained from \eqref{meanspincov} as
\be\label{meanspind}
S^\mu_{\Lambda*}({\bf p_*}) = [p_*]^\mu_\nu D^{S_\Lambda}({\sf J}^\nu)_{\lambda_\Lambda}^{\lambda'_\Lambda}
\Theta(\varphi_*,\theta_*)_{\lambda'_\Lambda \lambda_X}^{\lambda_\Lambda \lambda_X},
\ee
where the summation over $\lambda_X$ reduces the two-body SDM to the single-particle one for the Daughter. In general, the MSV of the Daughter depends on $(2j+1)^2-1$ real parameters through the $(2j+1)$-dimensional and trace $1$ hermitian SDM of the Mother. This means the MSV of the Daughter can not be definitely determined by the MSV of the Mother, which only involves $3$ real parameters, except for $j=1/2$. Indeed, this was well known in the literatures \cite{Leader,Kim:1992az} and illuminated explicitly in \cite{Xia:2019fjf}. Nevertheless, the SDM of the primary Mother can be well approximated by the first-order expansion form \eqref{spindensexp}, which surprisingly implies that the MSV of the Daughter can be definitely determined by that of the Mother now, as we will see in \eqref{meanspind2}.

By applying the approximation \eqref{spindensexp} for $\Theta(p_H)^m_{n}$ to the \eqref{spindensmat}, we explore the feed-down effect to first order in thermal vorticity $\varpi_*(p_H)$. The first term in \eqref{spindensexp} is proportional to the identity matrix and selects $m=n$ in \eqref{spindensmat}, then one is left with:
$$
D^j(\varphi_*,\theta_*,0)^{m \, *}_{\lambda} D^j(\varphi_*,\theta_*,0)^{m}_{\lambda'}
= \delta^\lambda_{\lambda'}
$$
due to the unitary of $D^j$'s. On the other hand, the second term gives 
rise to the three $D$-matrices multiplying term:
$$
D^j(\varphi_*,\theta_*,0)^{m \, *}_{\lambda} D^j({\sf J}^\rho)^m_n
D^j(\varphi_*,\theta_*,0)^{n}_{\lambda'} =D^{j\,(-1)}(\varphi_*,\theta_*,0)^{\lambda}_{m} 
D^j({\sf J}^\rho)^m_n D^j(\varphi_*,\theta_*,0)^{n}_{\lambda'},
$$
which, according to a well known relation in group representation theory~\cite{Tung}, equals to
\be\label{rotred}
{\sf R}(\varphi_*,\theta_*,0)^\rho_\tau  D^j({\sf J}^\tau)^\lambda_{\lambda'},
\ee
where the rotation ${\sf R}$ transforms the $z$ axis unit vector ${\bf e}_3$ into the ${\bf p}_*$ direction. Altogether, we get the explicit form of \eqref{spindensmat} as:
\be\label{spindensmat2}
\Theta(\varphi_*,\theta_*)_{\lambda'_\Lambda \lambda_X'}^{\lambda_\Lambda \lambda_X} \simeq
\frac{\delta^{\lambda}_{\lambda'} + {1\over2} \varpi_*(p_H)^{\alpha\beta} \epsilon_{\alpha\beta\rho\nu} 
	D^j({\sf J}^\tau)^\lambda_{\lambda'} {\sf R}(\varphi_*,\theta_*,0)^\rho_\tau {\hat t}^\nu}
{\left[T^j(\lambda_\Lambda,\lambda_X) T^j(\lambda'_\Lambda,\lambda'_X)^*\right]^{-1}\sum_{\lambda_\Lambda}^{\lambda_X} |T^j(\lambda_\Lambda,\lambda_X)|^2},
\ee
where the denominator gives the normalization factor solely determined by the dynamical amplitudes. For parity conservative decays with the property \eqref{phase2},  by substituting \eqref{spindensmat2} into \eqref{meanspind}, we find that the first term of \eqref{spindensmat2} does not contribute  as $\tr D^{S_\Lambda}({\sf J}^\nu)=0$ and the MSV of the Daughter is proportional to the thermal vorticity $\varpi_*(p_H)$ in the MRF:
\bea\label{meanspind2}
S^\mu_{\Lambda*}({\bf p_*}) & = &\frac{1}{2} \varpi_*(p_H)^{\alpha\beta} \epsilon_{\alpha\beta\rho\nu} {\hat t}^\nu 
\sum_{\lambda_X,\lambda'_X}\frac{\delta_{\lambda_X\lambda'_X}[p_*]^\mu_\kappa D^{S_\Lambda}({\sf J}^\kappa)_{\lambda_\Lambda}^{\lambda'_\Lambda} D^j({\sf J}^\tau)^\lambda_{\lambda'} 
	{\sf R}(\varphi_*,\theta_*,0)^\rho_\tau}{\left[T^j(\lambda_\Lambda,\lambda_X) T^j(\lambda'_\Lambda,\lambda'_X)^*\right]^{-1}\sum_{\lambda_\Lambda}^{\lambda_X} |T^j(\lambda_\Lambda,\lambda_X)|^2}
\nonumber \\
& = &- \frac{3S_{H*\rho}(p_H)}{j(j+1)}
\sum_{\lambda_X,\lambda'_X}\frac{\delta_{\lambda_X\lambda'_X}[p_*]^\mu_\kappa  D^{S_\Lambda}({\sf J}^\kappa)_{\lambda_\Lambda}^{\lambda'_\Lambda} D^j({\sf J}^\tau)^\lambda_{\lambda'} 
	{\sf R}(\varphi_*,\theta_*,0)^\rho_\tau}{\left[T^j(\lambda_\Lambda,\lambda_X) T^j(\lambda'_\Lambda,\lambda'_X)^*\right]^{-1}\sum_{\lambda_\Lambda}^{\lambda_X} |T^j(\lambda_\Lambda,\lambda_X)|^2}.
\eea
In the last step,  \eqref{vortspin} has been used to reexpress the formula in term of the MSV of the Mother in its rest
frame, $S_{H*}(p_H)$.  

As a first step, we would like to apply \eqref{meanspind2} to the simplest parity conservative decays: strong decays with the Byproduct $X=\pi$ and EM decays with $X=\gamma$. As mentioned in Sec.~\ref{Meanpolarization}, the dynamical amplitude has a definite sign under parity inversion in these cases, see \eqref{phase}. After that, the parity violating weak decays will be discussed in more detail with the Byproduct $X=\pi$.

\paragraph{\textbf{A. Strong decays}}

For the strong decay $H\rightarrow\Lambda+\pi$, the spin-parity structure is explicitly $j^{\eta_H}\rightarrow {1/2}^++0^-$. Hence,  \eqref{phase} becomes
$$T^j(-\lambda_\Lambda,0)=P_ST^j(\lambda_\Lambda,0), \qquad P_S\equiv\eta_H(-1)^{j+{1\over2}},$$
and \eqref{meanspind2} can be reduced to
\begin{align}\label{spin_strong}
S^\mu_{\Lambda*}({\bf p_*}) 
&= - \frac{3S_{H*\rho}(p_H)}{j(j+1)}
\frac{[p_*]^\mu_\kappa  D^{1/2}({\sf J}^\kappa)_{\lambda_\Lambda}^{\lambda'_\Lambda} D^j({\sf J}^\tau)^{\lambda_\Lambda}_{\lambda'_\Lambda}
	{\sf R}(\varphi_*,\theta_*,0)^\rho_\tau}{\left[T^j(\lambda_\Lambda,0) T^j(\lambda'_\Lambda,0)^*\right]^{-1}\sum_{\lambda_\Lambda=\pm{1\over2}} |T^j(\lambda_\Lambda,0)|^2}\nonumber\\
&=  - \frac{3S_{H*\rho}(p_H)}{j(j+1)}{P_S+(1-P_S)\delta_{\lambda_\Lambda\lambda'_\Lambda}\over2}
{[p_*]^\mu_\kappa  D^{1/2}({\sf J}^\kappa)_{\lambda_\Lambda}^{\lambda'_\Lambda} D^j({\sf J}^\tau)^{\lambda_\Lambda}_{\lambda'_\Lambda}
	{\sf R}(\varphi_*,\theta_*,0)^\rho_\tau}\nonumber\\
&=  \frac{3S_{H*\rho}(p_H)}{j(j+1)}\left\{
P_S C^j_\tau [p_*]^\mu_\tau
{\sf R}(\varphi_*,\theta_*,0)^{\rho\tau}+(1-P_S)C^j_3{[p_*]^\mu_3{\sf R}(\varphi_*,\theta_*,0)^{\rho3}}\right\}\nonumber\\
&=\frac{3S_{H*\rho}(p_H)}{j(j+1)}\left\{
P_S  C^j [p_*]^\mu_\tau{\sf R}(\varphi_*,\theta_*,0)^{\rho\tau} -(P_S  C^j-C^j_3){[p_*]^\mu_3{\sf R}(\varphi_*,\theta_*,0)^{\rho3}}\right\}.
\end{align}
In the derivation, the following conventions of $D^j(J)$ matrices are used for the Mother and Daughter:
\bea
\left\{\begin{array}{l}
	D^{1/2}({\sf J}^1)= {\sigma_1\over2}, \qquad D^{1/2}({\sf J}^2) = {\sigma_2\over2},
	\qquad D^{1/2}({\sf J}^3) = {\sigma_3\over2};\\
D^{3/2}_{\rm r}({\sf J}^1) = \sigma_1, \qquad\, D^{3/2}_{\rm r}({\sf J}^2) = \sigma_2,
\qquad D^{3/2}_{\rm r}({\sf J}^3) ={\sigma_3\over2};\end{array}\right.
\eea
where $D^{3/2}_{\rm r}$ are the $2\times2$ matrices for the Mother with spin $3/2$, reduced due to the restrictions of the indices $\lambda_\Lambda,\lambda'_\Lambda=\pm 1/2$. One can easily check that 
$$C_\tau^{1/2}=C^{1/2},\qquad C_\tau^{3/2} = \left({1\over2},{1\over2},{1\over2},{1\over4}\right) = C^{3/2}  - \frac{1}{4} \delta_\tau^3$$
with $C^{1/2}=1/4$ and $C^{3/2}=1/2$, where the irrelevant coefficients $C_0^j$, as ${\sf R}^{\rho0}=\eta^{\rho0}$, are introduced for the brevity of presentations.

In the helicity scheme, the matrix $[p_*]$ can be expanded according to \eqref{SLT}, so we take advantage of the orthogonality of rotations ${\sf R}$ to obtain
\begin{align}\label{spin_strong1}
S^\mu_{\Lambda*}({\bf p_*})&=\frac{3}{j(j+1)}\Big\{
P_S  C^j L_{\bf \hat p_*}(\xi)^\mu_\rho S_{H*}^\rho(p_H) -(P_S  C^j-C^j_3)S_{H*\rho}(P)
{\sf R}(\varphi_*,\theta_*,0)^\mu_\nu \nonumber\\
&\qquad\qquad\quad{\sf L}_z (\xi)^\nu_3 {\sf R}(\varphi_*,\theta_*,0)^{\rho 3}\Big\},
\end{align}
where $L_{\bf \hat p_*}(\xi)={\sf R}(\varphi_*,\theta_*,0){\sf L}_z (\xi){\sf R}^{-1}(\varphi_*,\theta_*,0)$ is the pure Lorentz boost transforming $\hat t$ into the ${\bf p_*}$ direction in the MRF. The Lorentz transformation involved in the second term of \eqref{spin_strong1} can be expressed explicitly as a function of ${\bf \hat p_*}$, that is,
\begin{align}\label{LT1}
&{\sf R}(\varphi_*,\theta_*,0)^\mu_\nu {\sf L}_z (\xi)^\nu_3 {\sf R}(\varphi_*,\theta_*,0)^{\rho 3}\nonumber\\
=& {\sf R}(\varphi_*,\theta_*,0)^\mu_3 {\sf L}_z (\xi)^3_3 {\sf R}(\varphi_*,\theta_*,0)^{\rho 3}
+ {\sf R}(\varphi_*,\theta_*,0)^\mu_0 {\sf L}_z (\xi)^0_3 {\sf R}(\varphi_*,\theta_*,0)^{\rho 3}
\nonumber \\
= & - \cosh \xi \, {\bf \hat p_*}^\mu  {\bf \hat p_*}^\rho - \sinh \xi \, {\bf \hat p_*}^\rho \delta^\mu_0
= - \frac{\varepsilon_{\Lambda*}}{m_{\Lambda}}{\bf \hat p_*}^\rho {\bf \hat p_*}^\mu - 
\frac{\p_*}{m_{\Lambda}} {\bf \hat p_*}^\rho \delta^\mu_0,
\end{align}
then \eqref{spin_strong1}  becomes
\bea
S^\mu_{\Lambda*}({\bf p_*})& =& \frac{3}{j(j+1)} \Big[ P_S  C^j L_{\bf \hat p_*}(\xi)^\mu_\rho S_{H*}^\rho(p_H)-(P_S  C^j-C^j_3)\Big(\frac{\varepsilon_{\Lambda*}}{m_\Lambda} {\bf S}_{H*}(p_H) \cdot {\bf \hat p_*} {\bf \hat p_*}^\mu \nonumber\\
&&\qquad \qquad+ \frac{\p_*}{m_\Lambda} {\bf S}_{H*}(p_H) \cdot {\bf \hat p_*} \delta^\mu_0\Big) \Big].
\eea
So with the help of the well known formulae for the pure Lorentz boost:
\bea\label{LT2}
L_{\bf \hat p_*}(\xi)^0_\rho={\varepsilon_{\Lambda*}\over m_\Lambda}\eta^0_\rho-{{\bf p}_{*\rho}\over m_\Lambda},\qquad L_{\bf \hat p_*}(\xi)^i_\rho=\eta^i_\rho-{{\bf p}_{*}^i{\bf p}_{*\rho}\over m_\Lambda(\varepsilon_{\Lambda*}+m_\Lambda)}-{{\bf p}_{*}^i\over m_\Lambda}\eta_{\rho0},
\eea
we get the explicit forms for the time and spatial components of the MSV of the Daughter in the MRF as
\begin{align*}
S^0_{\Lambda*}({\bf p_*}) & = 
\frac{3C^j_3}{j(j+1)} \frac{1}{m_\Lambda}  {\bf S}_{H*}(p_H)\cdot {\bf p}_*, \\
{\bf S}_{\Lambda*}({\bf p_*})
&= \frac{3}{j(j+1)} \left[P_S C^j{\bf S}_{H*}(p_H)-\left(P_S C^j-\frac{\varepsilon_{\Lambda*}}{m_\Lambda}C^j_3\right)
{\bf S}_{H*}(p_H) \cdot {\bf \hat p_*} 
{\bf \hat p_*}\right]. 
\end{align*}
Finally, we boost the MSV to the DRF as that is the one measured in experiments and find
\bea\label{lambdapol}
{\bf S}_{\Lambda o}({\bf p_*}) &=& {\bf S}_{\Lambda*}({\bf p_*}) - S^0_{\Lambda*}({\bf p_*}) \frac{\bf p_*}{\varepsilon_{\Lambda*}+m_\Lambda}\nonumber\\
&=& \frac{3}{j(j+1)} \left[P_S C^j{\bf S}_{H*}(p_H) - (P_S  C^j-C^j_3) {\bf S}_{H*}(p_H) \cdot {\bf \hat p_*} 
{\bf \hat p_*} \right].
\eea
The average over the whole solid angle $\Omega_*$ gives 
\bea\label{polInt1}
\langle{\bf S}_{\Lambda o}({\bf p_*})\rangle &=&\frac{3}{j(j+1)} \left[P_S C^j -  {1\over2}(P_S  C^j-C^j_3)\int_{0}^\pi \di\theta_*~\sin\theta_*\cos^2\theta_*\right]\langle{\bf S}_{H*}(p_H)\rangle\nonumber\\
& =&\frac{2P_S C^j+C^j_3}{j(j+1)}\langle{\bf S}_{H*}(p_H)\rangle
\eea
for a given ${\bf S}_{H*}(p_H)$ independent of ${\bf \hat p_*}$. The result is consistent with that found in Sec.~\ref{Meanpolarization}. In Sec.\ref{Momentumaverage}, we will see that ${\bf S}_{H*}(p_H)$ does depend on ${\bf \hat p_*}$ for a given momentum of the Daughter in the QGPF, thus the application of \eqref{polInt1} should be taken cautiously.

\paragraph{\textbf{B. Electromagnetic decays}}
For the EM decay $H\rightarrow\Lambda+\gamma$, the spin-parity structure is explicitly $j^{\eta_H}\rightarrow {1/2}^++1^-$. Hence, \eqref{phase} becomes 
$$T^j(-\lambda_\Lambda,-\lambda_X)=P_{EM}T^j(\lambda_\Lambda,\lambda_X), \qquad P_{EM}\equiv\eta_H(-1)^{j-{1\over2}}$$
with $ P_{EM}=-P_s$ for the same $\eta_H$ and $j$. In this case, \eqref{meanspind2} can be reduced to
\begin{align}
S^\mu_{\Lambda*}({\bf p_*}) 
&= - \frac{3S_{H*\rho}(p_H)}{j(j+1)}
\sum_{\lambda_X=\pm 1}^{\lambda'_X=\pm 1}\frac{\delta_{\lambda_X\lambda'_X}[p_*]^\mu_\kappa  D^{1/2}({\sf J}^\kappa)_{\lambda_\Lambda}^{\lambda'_\Lambda} D^j({\sf J}^\tau)^\lambda_{\lambda'} 
	{\sf R}(\varphi_*,\theta_*,0)^\rho_\tau}{\left[T^j(\lambda_\Lambda,\lambda_X) T^j(\lambda'_\Lambda,\lambda'_X)^*\right]^{-1}\sum^{\lambda_\Lambda=\pm 1/2}_{\lambda_X=\pm 1} |T^j(\lambda_\Lambda,\lambda_X)|^2}.
\end{align}
Keeping in mind the parity inversion properties of the irreducible representation matrices:
$$D^j({\sf J}^\tau)^{-\lambda}_{-\lambda'}=(-1)^{\delta_{\tau1}+1}D^j({\sf J}^\tau)^{\lambda}_{\lambda'}$$ 
with the component indices $|\lambda|,|\lambda'|\le j$, it is easy to show that 
\begin{align}\label{spin_EM}
S^\mu_{\Lambda*}({\bf p_*}) 
&= - \frac{3S_{H*\rho}(p_H)}{2j(j+1)}
\sum_{\lambda_X=\pm 1}^{\lambda'_X=\pm 1}\frac{\delta_{\lambda_X\lambda'_X}[p_*]^\mu_\kappa  \sum_{s=\pm}\left(D^{1/2}({\sf J}^\kappa)_{s\lambda_\Lambda}^{s\lambda'_\Lambda} D^j({\sf J}^\tau)^{s\lambda}_{s\lambda'}\right) 
	{\sf R}(\varphi_*,\theta_*,0)^\rho_\tau}{\left[T^j(\lambda_\Lambda,\lambda_X) T^j(\lambda'_\Lambda,\lambda'_X)^*\right]^{-1}\sum^{\lambda_\Lambda=\pm 1/2}_{\lambda_X=\pm 1} |T^j(\lambda_\Lambda,\lambda_X)|^2}\nonumber\\
&=-\frac{3S_{H*\rho}(p_H)}{j(j+1)}
\sum_{\lambda_X=\pm 1}^{\lambda'_X=\pm 1}\frac{\delta_{\lambda_X\lambda'_X}[p_*]^\mu_\kappa  D^{1/2}({\sf J}^\kappa)_{\lambda_\Lambda}^{\lambda'_\Lambda} D^j({\sf J}^\kappa)^{\lambda}_{\lambda'}
	{\sf R}(\varphi_*,\theta_*,0)^\rho_\kappa}{\left[T^j(\lambda_\Lambda,\lambda_X) T^j(\lambda'_\Lambda,\lambda'_X)^*\right]^{-1}\sum^{\lambda_\Lambda=\pm 1/2}_{\lambda_X=\pm 1} |T^j(\lambda_\Lambda,\lambda_X)|^2}\nonumber\\
&=-\frac{3S_{H*\rho}(p_H)}{j(j+1)}
\left(N^{j}-(N^{j}-N^{j}_3)\delta^{\kappa 3}\right){[p_*]^\mu_\kappa {\sf R}(\varphi_*,\theta_*,0)^\rho_\kappa},
\end{align}
where the $D^j({\sf J}^\kappa)$ and $T^j(\lambda_\Lambda,\lambda_X)$ relevant normalization factors are
\be\label{NF}
N^{j}= {\sqrt{3}\over2}{\sum_{\lambda_\Lambda=\pm 1/2}T^j(\lambda_\Lambda,1) T^j(-\lambda_\Lambda,1)^*\over\sum^{\lambda_\Lambda=\pm 1/2}_{\lambda_X=\pm 1}|T^j(\lambda_\Lambda,\lambda_X)|^2}, \ N^{j}_3=\!\!\sum_{\lambda_\Lambda=\pm 1/2}\!\!\!{(-1)^{\lambda_\Lambda+{1\over2}}(1\!-\!\lambda_\Lambda)|T^j(\lambda_\Lambda,1)|^2\over\sum^{\lambda_\Lambda=\pm 1/2}_{\lambda_X=\pm 1}|T^j(\lambda_\Lambda,\lambda_X)|^2}.
\ee
Now, we can immediately identify the similarity between \eqref{spin_strong} and \eqref{spin_EM}. So the final results for the MSV of the Daughter and the averaged one in the DRF can be given directly as
\bea\label{lambdapol2}
{\bf S}_{\Lambda o}({\bf p_*}) &=& \frac{3}{j(j+1)} \left[N^{j}{\bf S}_{H*}(p_H) - (N^{j}-N^{j}_3) {\bf S}_{H*}(p_H) \cdot {\bf \hat p_*}
{\bf \hat p_*} \right],\\
\langle{\bf S}_{\Lambda o}({\bf p_*})\rangle 
& =&\frac{2N^{j}+N^{j}_3}{j(j+1)}\langle{\bf S}_{H*}(p_H)\rangle
\eea
by changing the coefficients 
$$P_S C^j\rightarrow N^{j},\qquad\qquad C^j_3\rightarrow N^{j}_3$$
from \eqref{lambdapol} and \eqref{polInt1}.

For $j=3/2$, the formula \eqref{lambdapol2} can not be simplified further in general, as the transition amplitudes in \eqref{NF} can not be canceled out. However, fortunately for the study of $\Lambda$ polarization, only the EM decay $\Sigma^0 \rightarrow \Lambda \gamma$ is relevant and $j^{\eta_{\Sigma^0}}={1/2^+}$. Then, we immediately find that $N^{1/2}=0$ and $N^{1/2}_3=-1/4$ due to the restrictions $|\lambda|,|\lambda'|\leq 1/2$~\cite{Becattini:2016gvu}, so the MSV and the averaged one are explicitly
\bea\label{lambdapol3}
{\bf S}_{\Lambda o}({\bf p_*}) &=& -{\bf S}_{H*}(p_H) \cdot {\bf \hat p_*} \, {\bf \hat p_*},\\
\langle{\bf S}_{\Lambda o}({\bf p_*})\rangle  &=&-{1\over3}\langle{\bf S}_{H*}(p_H)\rangle.
\eea
The result is also consistent with that found in Sec.~\ref{Meanpolarization}. 

\paragraph{\textbf{C. Weak decays}}
For weak decays, it is well known that the dynamical transition amplitude is a mixture of parity even and odd modes. Only one kind of weak decay channel is relevant to $\Lambda$ polarization, that is, $\Xi\rightarrow\Lambda+\pi$, so we stick to the simple case with $j^{\eta_{\Xi}}={1/2^+}$. First of all, due to parity violation, the first term of \eqref{spindensmat2} will give rise to a finite contribution to the MSV of the Daughter~\cite{Xia:2019fjf}. Assuming the dynamical amplitude in the following form
$$
T^{1/2}_w(\pm 1/2,0)={T_e\pm T_o},
$$
this contribution is simply
\bea\label{PV}
{S_{\Lambda*}'}^{\mu}({\bf p_*})={\alpha_w\over 2m_\Lambda} (p_*\eta^{\mu0}+\varepsilon_{\Lambda*}{\bf \hat p_*}^\mu),\qquad \alpha_w={2\Re(T^{*}_e T_o)\over\left|T_e\right|^2+\left|T_o\right|^2},
\eea
and the corresponding MSV in the DRF is proportional to the three-momentum unit vector:
\bea
{\bf S}_{\Lambda o}'({\bf p_*})={\alpha_w\over 2}{\bf \hat p_*}.
\eea
Next, the polarization transfer effect from the Mother can be deduced from \eqref{meanspind2} as
\begin{align}\label{meanspind3}
{S_{\Lambda*}''}^{\mu}({\bf p_*}) 
&= - {4S_{H*\rho}(p_H)}
\frac{[p_*]^\mu_\kappa  D^{1/2}({\sf J}^\kappa)_{\lambda_\Lambda}^{\lambda'_\Lambda} D^{1/2}({\sf J}^\tau)^{\lambda_\Lambda}_{\lambda'_\Lambda}
	{\sf R}(\varphi_*,\theta_*,0)^\rho_\tau}{\left[T^j(\lambda_\Lambda,0) T^j(\lambda'_\Lambda,0)^*\right]^{-1}\sum_{\lambda_\Lambda=\pm{1\over2}} |T^j(\lambda_\Lambda,0)|^2}\nonumber\\
&= - S_{H*\rho}(p_H)
{[p_*]^\mu_\kappa \left((1-\gamma_w)\delta^{\kappa3}\delta^{\tau3}-\gamma_w\eta^{\kappa\tau}+\epsilon^{\kappa\tau 3}\beta_w\right)
	{\sf R}(\varphi_*,\theta_*,0)^\rho_\tau}
\end{align}
with the dynamical parameters
\bea\label{bega}
\beta_w={2\Im(T^{*}_e T_o)\over\left|T_e\right|^2+\left|T_o\right|^2},\qquad\gamma_w={\left|T_e\right|^2-\left|T_o\right|^2\over\left|T_e\right|^2+\left|T_o\right|^2}.
\eea

Again, we can immediately recognize the similarity between the first two terms of \eqref{meanspind3} and those in the strong decay \eqref{spin_strong}, hence simple alternations of the coefficients will give the final results. By noticing $L_{z}(\xi)^\nu_{\tau}=\delta^\nu_{\tau}$ for $\tau=1,2$ in \eqref{standard}, the Lorentz transformation in the last term of \eqref{meanspind3} can be evaluated as
$$
\epsilon^{\kappa\tau 3}[p_*]^\mu_\kappa	{\sf R}(\varphi_*,\theta_*,0)^\rho_\tau=\epsilon^{\kappa\tau 3}{\sf R}(\varphi_*,\theta_*,0)^\mu_\kappa	{\sf R}(\varphi_*,\theta_*,0)^\rho_\tau.
$$
As we already know the explicit forms of the involved rotations:
\bea
{\sf R}(\varphi_*,\theta_*,0)^{\mu}_1&=&(\cos\varphi_*\cos\theta_*,\sin\varphi_*\cos\theta_*,-\sin\theta_*),\nonumber\\
{\sf R}(\varphi_*,\theta_*,0)^{\mu}_2&=&(-\sin\varphi_*,\cos\varphi_*,0),\nonumber
\eea
the transformation can be shown to be simply
\bea\label{LT3}
\epsilon^{\kappa\tau 3}{\sf R}(\varphi_*,\theta_*,0)^\mu_\kappa	{\sf R}(\varphi_*,\theta_*,0)^\rho_\tau=\epsilon^{\mu\nu\rho} {\bf\hat{p}_{*\nu}}.
\eea
So, gathering \eqref{LT1}, \eqref{LT2} and \eqref{LT3} all in \eqref{meanspind3}, we will find
\begin{align*}
{S_{\Lambda*}''}^{0}({\bf p_*}) & = 
\frac{1}{m_\Lambda}  {\bf S}_{H*}(p_H)\cdot {\bf p}_*, \\
{\bf S}_{\Lambda*}''({\bf p_*})
&=\gamma_w{\bf S}_{H*}(p_H)\!+\!\frac{\varepsilon_{\Lambda*}\!-\!\gamma_wm_\Lambda}{m_\Lambda}
{\bf S}_{H*}(p_H) \cdot {\bf \hat p_*} 
{\bf \hat p_*}\!+\!\beta_w{\bf S}_{H*}(p_H) \times {\bf \hat p_*},
\end{align*}
and the MSV in the DRF is
\bea\label{lambdapol4}
{\bf S}_{\Lambda o}''({\bf p_*})
&=& \gamma_w{\bf S}_{H*}(p_H)+(1-\gamma_w)
{\bf S}_{H*}(p_H) \cdot {\bf \hat p_*} 
{\bf \hat p_*}+\beta_w{\bf S}_{H*}(p_H) \times {\bf \hat p_*}\nonumber\\
&=& {\bf S}_{H*}(p_H) \cdot {\bf \hat p_*} 
{\bf \hat p_*}+\beta_w{\bf S}_{H*}(p_H) \times {\bf \hat p_*}+\gamma_w{\bf \hat p_*}\times({\bf S}_{H*}(p_H) \times {\bf \hat p_*}).
\eea
Finally, the total MSV of the Daughter is 
$${\bf S}_{\Lambda o}({\bf p_*})={\bf S}_{\Lambda o}'({\bf p_*})+{\bf S}_{\Lambda o}''({\bf p_*}),$$
and the average over the whole solid angle $\Omega_*$ gives
\bea
\langle{\bf S}_{\Lambda o}({\bf p_*})\rangle& =&\left[\gamma_w + (1-\gamma_w){1\over2}\int_{0}^\pi \di\theta_*~\sin\theta_*\cos^2\theta_*\right]\langle{\bf S}_{H*}(p_H)\rangle\nonumber\\
&=&{1+2\gamma_w\over3}\langle{\bf S}_{H*}(p_H)\rangle.
\eea
As expected from the arguments in Sec.\ref{Meanpolarization}, the spontaneous local polarization ${\bf S}_{\Lambda o}'({\bf p_*})$ doesn't contribute to the global one.

For the convenience of future use, we summary all the polarization transfer from the decays of the Mother with polarization vector ${\bf P}_{H*}={\bf S}_{H*}(p_H)/j$ to the Daughter with polarization vector ${\bf P}_{\Lambda}=2{\bf S}_{\Lambda o}({\bf p_*})$ in Table.\ref{poltran}, where explicit decay channels are also listed. The results are completely consistent with those given in Ref.\cite{Xia:2019fjf} to linear order of the thermal vorticity. We notice that ${P}_S=-1$ in the strong decay ${1/2^+}\rightarrow{1/2^+}0^-$, that is, only the dynamical amplitude with odd parity is involved. Thus, the polarization transfer result can be alternatively derived from the more general formula of the weak decay ${1/2^+}\rightarrow{1/2^+}0^-$ by setting $\alpha_w=\beta_w=0$ and $\gamma_w=-1$ as $T_e=0$. The results are truly consistent with each other according to Table.\ref{poltran}.
\begin{table}
	\caption{Polarization transfer formulae for the decay $H\rightarrow\Lambda+X$ in the Mother's rest frame.}\label{poltran}
	\begin{tabular}{lcc}
		\hline\noalign{\smallskip}
		Decay channels & Local polarization ${\bf P}_{\Lambda}$ &$\langle{\bf P}_{\Lambda}\rangle/\langle{\bf P}_{H*}\rangle$  \\
		\noalign{\smallskip}\svhline\noalign{\smallskip}
		A. Strong$^a$ & $\frac{6}{(j+1)} \left[P_S C^j{\bf P}_{H*}- (P_S  C^j-C^j_3) {\bf P}_{H*} \cdot {\bf \hat p_*} 
		{\bf \hat p_*} \right]$  & $\frac{2(2P_S C^j+C^j_3)}{(j+1)}$\\
		\quad${1/2^+}\rightarrow{1/2^+}0^-$ &  $-{\bf P}_{H*}+2{\bf P}_{H*} \cdot {\bf \hat p_*} 
		{\bf \hat p_*}$ & $-{1/3}$\\
		\quad${1/2^-}\rightarrow{1/2^+}0^-$ &  ${\bf P}_{H*}$  & $1$\\
		\quad${3/2^+}\rightarrow{1/2^+}0^-$ & ${3\over5}\left[2{\bf P}_{H*}-{\bf P}_{H*} \cdot {\bf \hat p_*} 
		{\bf \hat p_*}\right]$  & $1$\\
		\quad${3/2^-}\rightarrow{1/2^+}0^-$ &  ${3\over5}\left[-2{\bf P}_{H*}+3{\bf P}_{H*} \cdot {\bf \hat p_*} 
		{\bf \hat p_*}\right]$ & $-{3/5}$\\
		\noalign{\smallskip}\hline\noalign{\smallskip}
		B. Electromagnetic$^b$ & $\frac{6}{(j+1)} \left[N^j{\bf P}_{H*}- (N^j-N^j_3) {\bf P}_{H*} \cdot {\bf \hat p_*} 
		{\bf \hat p_*} \right]$  & $\frac{2(2N^j+N^j_3)}{(j+1)}$\\
		\quad${1/2^\pm}\rightarrow{1/2^+}1^-$ &  $-{\bf P}_{H*} \cdot {\bf \hat p_*} \, {\bf \hat p_*}$  & $-{1/3}$\\
		\noalign{\smallskip}\hline\noalign{\smallskip}
		C. Weak$^c$ & & \\
		\quad${1/2^+}\rightarrow{1/2^+}0^-$ & $(\alpha_w+{\bf P}_{H*} \cdot {\bf \hat p_*}) 
		{\bf \hat p_*}+\beta_w{\bf P}_{H*} \times {\bf \hat p_*}+\gamma_w{\bf \hat p_*}\times({\bf P}_{H*}\times {\bf \hat p_*})$  & ${1+2\gamma_w\over3}$\\
		\noalign{\smallskip}\hline\noalign{\smallskip}
	\end{tabular}
\\$^a$ $P_S\equiv\eta_H(-1)^{j+{1\over2}}, C^{1/2}=1/4, C^{3/2}=1/2$ and $C_3^{1/2}=C_3^{3/2}=1/4.$
\\$^b$ See \eqref{NF} for the definitions of $N^j$ and $N^j_3$.
\\$^c$ See \eqref{PV} and \eqref{bega} for the definitions of $\alpha_w,\beta_w$ and $\gamma_w$.
\vspace*{-12pt}
\end{table}
\subsubsection{Average over the momentum of the Mother}\label{Momentumaverage}

\begin{figure}[t]
	\sidecaption[t]
  \includegraphics[scale=.2]{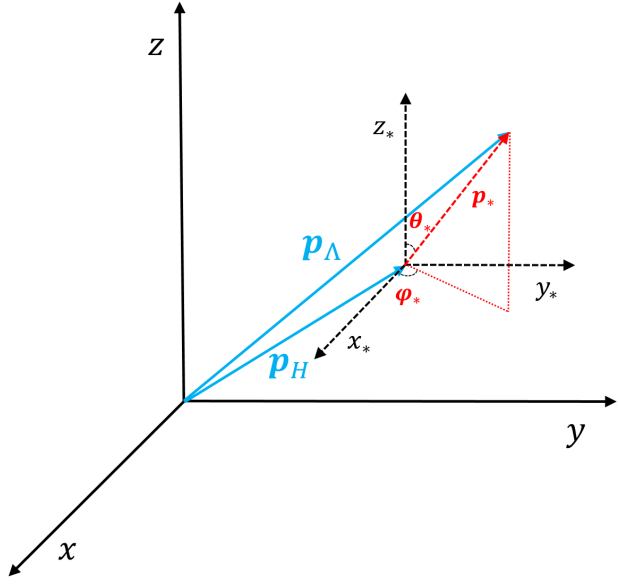}
	\caption{The coordinate systems in the QGP frame, with solid axis and vector lines, and the Mother's rest frame, with dashed axis and vector lines. ${\bf p}_H$ and ${\bf p}_\Lambda$ are the momenta of the Mother and the Daughter in the QGP frame, respectively. ${\bf p}_*$ is the momentum of the Daughter in the Mother's rest frame with the azimuthal angle $\varphi_*$ and polar angle $\theta_*$.}\label{CSs} 
\end{figure}

In previous section, we have established the formulae for the polarization transfer in two-body decays, where the momentum of the Daughter is given in the Mother's rest frame. However, we are more interested in the polarization inherited by the Daughter as a function of its momentum ${\bf p}_\Lambda$ in the QGP frame. In the QGPF, the Mother is in a momentum distribution which has to be averaged over before useful results are obtained to compare with experimental measurements. So first of all in this section, we establish the Mother's momentum averaged formula for the mean spin vector of the Daughter with a given momentum in the QGPF. The coordinate systems are parallel to each other in the QGPF and the MRF, see Fig.\ref{CSs}, where the momenta of the Mother ${\bf p}_H$ and the Daughter ${\bf p}_\Lambda$ in the QGPF and the momentum of the Daughter ${\bf p}_*$ in the MRF are also illuminated.
Note that these momenta are related to each other through the Lorentz boost from the QGPF to the MRF, rather than the simple triangle algebra for vectors in a single coordinate system, see Appendix.\ref{app:lorentz}.

Let $n({\bf p}_H)$ be the un-normalized momentum distribution of the Mother in the QGPF such that ${\int \di^3 {\rm p}_H \; n({\bf p}_H)}$ yields the total number of the Mother, one would then define the MSV of the Daughter fed-down from a specific decay as:
$$
 {\bf S}_{\Lambda o}({\bf p}_\Lambda)= 
\frac{\int \di^3 {\rm p}_H \; n({\bf p}_H) \, {\bf S}_{\Lambda o}({\bf p_*})}{\int \di^3 {\rm p}_H \; n({\bf p}_H)},
$$
where the MSV of the Daughter in the MRF ${\bf S}_{\Lambda o}({\bf p_*})$ is listed in the second column of Table.\ref{poltran}. Since the magnitude of ${\bf p}_*$ is fixed in two-body decay, see \eqref{pmagn}; the three components of ${\bf p}_H$ are not completely independent for a given ${\bf p}_\Lambda$, see the Lorentz boost relation:
$$
\varepsilon_{\Lambda*} = \frac{\varepsilon_H}{m_H} \varepsilon_\Lambda - \frac{1}{m_H} {\bf p}_H \cdot {\bf p}_\Lambda = \sqrt{\p^2_{\Lambda*} + m^2_\Lambda}
$$
with the energy $\varepsilon_{H/\Lambda}=\sqrt{{\bf p}_{H/\Lambda}^2+m_{H/\Lambda}^2}$ in the QGPF. Taking into account this fact, one should redefine the MSV of the Daughter with momentum ${\bf p}_\Lambda$ in the QGPF by multiplying the integrands by a delta function, that is,
\be\label{Mvar}
{\bf S}_{\Lambda o}({\bf p}_\Lambda) = 
\frac{\int \di^3 {\rm p}_H \; n({\bf p}_H) \, {\bf S}_{\Lambda o}({\bf p_*}) \delta(p_* - p_{\Lambda*})}
{\int \di^3 {\rm p}_H \; n({\bf p}_H) \delta(p_* - p_{\Lambda*})}.
\ee
By altering the integration variable from ${\bf p}_H$ to ${\bf p}_*$ through the Lorentz boost relation (see Appendix.1):
\be\label{momrel}
{\bf p}_H =  \frac{2 m_H(\varepsilon_{\Lambda*} + \varepsilon_\Lambda)({\bf p}_\Lambda - {\bf p_*})}
{(\varepsilon_{\Lambda*} + \varepsilon_\Lambda)^2-({\bf p}_\Lambda - {\bf p_*})^2} = 
 \frac{m_H(\varepsilon_{\Lambda*} + \varepsilon_\Lambda)({\bf p}_\Lambda - {\bf p_*})}
{m_\Lambda^2 + \varepsilon_\Lambda\varepsilon_{\Lambda*} +{\bf p}_\Lambda\cdot{\bf p_*}}
\implies 
\hat{\bf p}_H = \frac{{\bf p}_\Lambda - {\bf p_*}}{|{\bf p}_\Lambda - {\bf p_*}|}
\ee
and completing the integrations over the magnitude $p_*$, only solid angle integrations are left over:
\be\label{decayspin}
{\bf S}_{\Lambda o}({\bf p}_\Lambda) =  
\frac{\int \di \Omega_* \; n({\bf p}_H) \left\| \frac{\partial {\bf p}_H}{\partial{\bf p}_*} \right\|\, 
	{\bf S}_{\Lambda o}({\bf p}_*)}
{\int \di \Omega_* n({\bf p}_H)  \left\| \frac{\partial \bf P}{\partial {\bf p}_*} \right\| \,}.
\ee
Here and in the following, one should keep in mind that $p_*$ is fixed to $p_{\Lambda*}$ and the absolute value of the determinant of the Jacobian (AVDJ) reads (see Appendix.1):
\be\label{jacob}
\left\| \frac{\partial {\bf p}_H}{\partial{\bf p}_*} \right\|= \frac{m_H^3(\varepsilon_{\Lambda*} + \varepsilon_\Lambda)^2 \left[ (\varepsilon_{\Lambda*} + \varepsilon_\Lambda)^2 - (\varepsilon_\Lambda \varepsilon_{\Lambda*} + {\bf p}_\Lambda\cdot {\bf p}_*+m_\Lambda^2) \right]}{\varepsilon_{\Lambda*}(\varepsilon_\Lambda \varepsilon_{\Lambda*} + {\bf p}_\Lambda\cdot {\bf p}_*+m_\Lambda^2)^3}.
\ee

The most involved thing in the evaluation of \eqref{decayspin} is that ${\bf S}_{\Lambda o}({\bf p}_*)$ implicitly depends on $\varphi_*$ and $\theta_*$ through ${\bf S}_{H*}(p_H)$, besides explicitly through ${\bf \hat p_*}$. The features of ${\bf S}_{H*}(p_H)$ have been well studied by following the symmetries, associated with the parity inversion and rotation around the total angular momentum axis, of the fireball produced in peripheral heavy ion collisions. So the three components of ${\bf S}_{H*}(p_H)$ can be expanded as Fourier series of the momentum azimuthal angle $\varphi_H$ to the second-order harmonics~\cite{Becattini:2017gcx,Xia:2018tes,Becattini:2015ska}:
\begin{align}\label{harmonics}
S_{H*x} &\simeq \frac{2j(j+1)}{3} \left[ h_1({\rm p}_H^T,Y_H) \sin \varphi_H
+ h_2({\rm p}_H^T,Y_H) \sin 2 \varphi_H \right] ,  \nonumber \\
S_{H*y} &\simeq \frac{2j(j+1)}{3} \left[ g_0({\rm p}_H^T,Y_H) + g_1({\rm p}_H^T,Y_H) \cos \varphi_H 
+ g_2({\rm p}_H^T,Y_H) \cos 2 \varphi_H \right], \nonumber \\
S_{H*z} &\simeq \frac{2j(j+1)}{3} f_2({\rm p}_H^T,Y_H) \sin 2 \varphi_H,
\end{align}
where ${\rm p}_H^T$ and $Y_H$ are the magnitudes of the transverse momentum and the rapidity of the Mother, respectively. According to the \eqref{vortspin}, the prefactor ${2j(j+1)/3}$ is extracted out from all the functions $f,g$ and $h$ so that they don't depend on the total spin $j$ any more. The aforementioned symmetries imply that $h_1$ and
$g_1$ are odd functions of $Y_H$ whereas $g_0, f_2, g_2$ and $ h_2$ are even. Furthermore, in a right-handed reference frame with $x$-axis on the reaction plane and $y$-axis in the
direction opposite to the total angular momentum, both the hydrodynamic model~\cite{Karpenko:2016jyx} and AMPT model~\cite{Xia:2018tes} prediced the magnitudes of all the coefficient functions and particularly their signs to be:
\bea
&&h_1({\rm p}_H^T,Y_H>0)>0, \quad h_2({\rm p}_H^T,Y_H)< 0, \quad g_0({\rm p}_H^T,Y_H)< 0,\nonumber\\
&&g_1({\rm p}_H^T,Y_H>0)<0, \quad g_2({\rm p}_H^T,Y_H)> 0, \quad f_2({\rm p}_H^T,Y_H)< 0.\label{signs}
\eea
For the study of $\Lambda$ polarization, the Mother's masses are at most $24\%$ larger than that of $\Lambda$, so we can assume $f,g$ and $h$ to be the same as those for the primary $\Lambda$ according to \eqref{varpiboost}. 

It's more convenient to represent ${\bf p}_H$ and ${\bf p}_\Lambda$ with cylindrical coordinates and ${\bf p_*}$ with the spherical ones as:  
\begin{align}
{\bf p}_H ={}& {\rm p}_H^T \cos \varphi_H {\bf e}_1 + {\rm p}_H^T \sin \varphi_H {\bf e}_2+
{\rm p}_{Hz} {\bf e}_3,  \nonumber \\
{\bf p}_\Lambda={}& {\p_\Lambda^T} \cos \varphi_\Lambda {\bf e}_1 + {\p_\Lambda^T} \sin \varphi_\Lambda {\bf e}_2+ \p_{\Lambda z} {\bf e}_3,\nonumber \\
{\bf p_*}={}& \p_* \sin \theta_* \cos \varphi_* {\bf e}_1 + \p_* \sin \theta_* \sin \varphi_* {\bf e}_2
+ \p_* \cos \theta_* {\bf e}_3.
\end{align}
In the following, we stick to the simplest case of midrapidity $\Lambda$ with $\p_{\Lambda z} = 0$. . Then, the rightmost equality in \eqref{momrel} can be used to express the trigonometric functions of the Mother in terms of the spherical coordinates of the Daughter as:
\bea\label{phimother}
\sin 2 \varphi_H  & = & \frac{ \p_*^2 \sin^2\theta_* \sin 2 \varphi_* + {\p_\Lambda^T}^2 
	\sin 2\varphi_\Lambda - 2 \p_* {\p_\Lambda^T} \sin \theta_* \sin (\varphi_* +\varphi_\Lambda)}
{\p_*^2 \sin^2\theta_*+{\p_\Lambda^T}^2 - 2\p_* {\p_\Lambda^T} \sin \theta_* \cos (\varphi_* -\varphi_\Lambda)} \nonumber\\
&=& {\cal A}(\theta_*,\psi)\sin2\varphi_\Lambda +{\cal B}(\theta_*,\psi)\cos2\varphi_\Lambda, \nonumber\\
\cos 2 \varphi_H  &=& \frac{\p_*^2 \sin^2\theta_* \cos 2 \varphi_* + {\p_\Lambda^T}^2 \cos 2\varphi_\Lambda - 
	2 \p_* {\p_\Lambda^T} \sin \theta_* \cos(\varphi_*+\varphi_\Lambda)}{\p_*^2 \sin^2\theta_*+{\p_\Lambda^T}^2 - 2\p_* {\p_\Lambda^T} 
	\sin \theta_* \cos (\varphi_* -\varphi_\Lambda)}
\nonumber\\
&=& {\cal A}(\theta_*,\psi)\cos 2\varphi_\Lambda - {\cal B}(\theta_*,\psi)\sin2\varphi_\Lambda,  \nonumber \\
\sin \varphi_H  & = & \frac{{\p_\Lambda^T} 
	\sin \varphi_\Lambda - \p_* \sin\theta_* \sin \varphi_*}
{\sqrt{\p_*^2 \sin^2\theta_*+{\p_\Lambda^T}^2 - 2\p_* {\p_\Lambda^T} \sin \theta_* \cos (\varphi_* -\varphi_\Lambda)}}  \nonumber\\
&=& {\cal C}(\theta_*,\psi)\sin \varphi_\Lambda +{\cal D}(\theta_*,\psi)\cos \varphi_\Lambda, \nonumber\\
\cos \varphi_H  &=& \frac{{\p_\Lambda^T} 
	\cos \varphi_\Lambda - \p_* \sin\theta_* \cos \varphi_*}{\sqrt{\p_*^2 \sin^2\theta_*+{\p_\Lambda^T}^2 - 2\p_* 
		{\p_\Lambda^T} \sin \theta_* \cos (\varphi_* -\varphi_\Lambda)}}\nonumber\\
&=& {\cal C}(\theta_*,\psi)\cos \varphi_\Lambda - {\cal D}(\theta_*,\psi)\sin \varphi_\Lambda
\eea
with the introduced variable $\psi = \varphi_* - \varphi_\Lambda $ and the auxiliary functions:
\bea\label{coeff}
{\cal A}(\theta_*,\psi)&=&{\p_*^2\sin^2\theta_* \cos 2 \psi - 2 \p_* {\p_\Lambda^T} \sin\theta_* \cos\psi +{\p_\Lambda^T}^2 
	\over \p_*^2 \sin^2\theta_*+{\p_\Lambda^T}^2 - 2\p_* {\p_\Lambda^T} \sin \theta_* \cos \psi},\nonumber\\
{\cal B}(\theta_*,\psi)&=&{\p_*^2 \sin^2\theta_* \sin 2\psi - 2\p_* {\p_\Lambda^T} \sin\theta_* \sin\psi 
	\over \p_*^2 \sin^2\theta_*+{\p_\Lambda^T}^2 - 2\p_* {\p_\Lambda^T} \sin \theta_* \cos \psi}, \nonumber \\
{\cal C}(\theta_*,\psi)&=&{{\p_\Lambda^T} - \p_* \sin\theta_* \cos \psi 
	\over \sqrt{\p_*^2 \sin^2\theta_*+{\p_\Lambda^T}^2 - 2\p_* {\p_\Lambda^T} \sin \theta_* \cos \psi}},\nonumber\\
{\cal D}(\theta_*,\psi)&=&{- \p_* \sin\theta_* \sin \psi  
	\over \sqrt{\p_*^2 \sin^2\theta_*+{\p_\Lambda^T}^2 - 2\p_* {\p_\Lambda^T} \sin \theta_* \cos \psi}}.
\eea
One can easily verify the even-odd and normalization features of the auxiliary functions, that is,
\begin{align}
{\cal A}(\theta_*,-\psi) &= {\cal A}(\theta_*,\psi), \qquad \qquad {\cal B}(\theta_*,-\psi) 
= -{\cal B}(\theta_*,\psi), \nonumber\\
{\cal C}(\theta_*,-\psi) &= {\cal C}(\theta_*,\psi), \qquad \qquad \, {\cal D}(\theta_*,-\psi), 
= -{\cal D}(\theta_*,\psi)\nonumber\\
{\cal A}^2(\theta_*,\psi)+&{\cal B}^2(\theta_*,\psi)=1,\qquad \quad{\cal C}^2(\theta_*,\psi)+{\cal D}^2(\theta_*,\psi)=1.
\end{align}
With the variable transformation $\varphi_*\rightarrow\psi$, the integration over the solid angle $\di \Omega_*$ in \eqref{decayspin} can be replaced by another one:
\bea
\int \di \Omega_*= \int_0^\pi \di \theta_* \sin \theta_* \int^{2\pi-\varphi_\Lambda}_{-\varphi_\Lambda} \di \psi
= \int_0^\pi \di \theta_* \sin \theta_* \int_{-\pi}^{\pi} \di \psi, \nonumber
\eea
where the last step is owing to the $2\pi$-periodic in $\psi$ of all the functions in the integrands. 

The spectrum function $n({\bf p}_H)$ depends on the specific model of the collision, but it must be {\em even} 
in "$\cos \theta_H$" because of the symmetries of the colliding system and isotropic in the transverse plane when the usually small elliptic flow is neglected. So, $n({\bf p}_H)$ can be assumed to only depend on the magnitudes of its longitudinal and transverse momenta ${\rm p}_H^L$ and ${\rm p}_H^T$ to a very good approximation. In this case, ${\rm p}_H^L$ and ${\rm p}_H^T$ can be given explicitly with the variables for the Daughter by following \eqref{momrel} as:
\bea
{\rm p}_H^L &=& m_H \frac{(\varepsilon_{\Lambda*} + \varepsilon_\Lambda) 
	|\p_*\cos \theta_*|} 
{m_\Lambda^2 + \varepsilon \varepsilon_{\Lambda*} + {\p_\Lambda^T} \p_{*} \sin\theta_* \cos\psi},\label{plong}\\
{\rm p}_H^T &=& m_H \frac{(\varepsilon_{\Lambda*} + \varepsilon_\Lambda) 
	\sqrt{\p_*^2 \sin^2\theta_* + {\p_\Lambda^T}^2 - 2{\p_\Lambda^T} \p_* \sin\theta_* \cos\psi}} 
{m_\Lambda^2 + \varepsilon \varepsilon_{\Lambda*} + {\p_\Lambda^T} \p_{*} \sin\theta_* \cos\psi},\label{ptran}
\eea
which imply that the Mother's spectrum function is even in both $\cos \theta_*$ and $\psi$. Then, as the polar angle of the Mother is given by
\be
\cos \theta_H \equiv \hat{\bf p}_H \cdot {\bf e}_3= 
- \frac{\p_*\cos \theta_*}{|{\bf p}_\Lambda - {\bf p_*}|} =  - \frac{\p_*\cos \theta_*}{\sqrt{  
		\p_*^2+{\p_\Lambda^T}^2-2{\p_\Lambda^T} \p_* \sin\theta_* \cos\psi}},
\ee
the rapidity $Y_H$ is found to be an odd function of $\cos \theta_*$ and even function of 
$\psi$, that is,
$$
m_H^T \sinh Y_H = {\bf p}_H \cdot {\bf e}_3
= {\rm p}_H^T \tan \theta_H,
$$
where the transverse mass $m_H^T = \sqrt{({\rm p}_H^T)^2 + m_H^2}$.
For the convenience of future discussions, we summarize the even-oddness of all the functions relevant to the evaluation of \eqref{decayspin} in Table.\ref{evenodd}. 
\begin{table}
\centering
	\caption{The even ("$+$") and odd ("$-$") properties of relevant functions in \eqref{decayspin} (top row) with respect to the variables (first column).}\label{evenodd}
	\begin{tabular}{ccccccc}
		\hline\noalign{\smallskip}
		 $\ \ $ Variables $\ \ $ & $\ \ h_1,g_1\ \ $ & $\ \ h_2,g_0,g_2,f_2\ \ $ & $\ \ {\cal A},{\cal C}\ \ $ & $\ \ {\cal B},{\cal D}\ \ $ & $\ \ n({\bf p}_H)\ \ $ & $\ \ \left\| \frac{\partial {\bf p}_H}{\partial{\bf p}_*} \right\|\ \ $\\
		\noalign{\smallskip}\svhline\noalign{\smallskip}
		$\cos \theta_*$& $-$ & $+$ & $+$ & $+$ & $+$ & $+$\\
		\noalign{\smallskip}\hline\noalign{\smallskip}
		$\psi$ & $+$ & $+$ & $+$ & $-$ & $+$ & $+$\\
		\noalign{\smallskip}\hline\noalign{\smallskip}
	\end{tabular}
\end{table}

Now, we can well understand the advantage of introducing the new variable "$\psi$": In this way, ${\rm p}_H^T$ and ${\rm p}_{H}^L$, thus $g,f,h,n({\bf p}_H)$ and $\left\| \frac{\partial {\bf p}_H}{\partial{\bf p}_*} \right\|$, are all independent of the observable $\varphi_\Lambda$ and the integrations over the new solid angles $\psi$ and $\theta_*$ can be numerically carried out easily. 

We now pay attention to the parity-conservative strong and EM decays first, which share very similar expressions for the MSV of the Daughter, see Table.\ref{poltran}. For brevity, the following general formula will be used:
\be\label{SPC}
{\bf S}_{\Lambda o}^{PC}({\bf p}_*)=\frac{3}{j(j+1)} \left[A\,{\bf S}_{H*}+B\, {\bf S}_{H*} \cdot {\bf \hat p_*} 
{\bf \hat p_*} \right],\ee
where the strong (EM) decay coefficients $A=P_S  C^j\,(N^j)$ and $B=C^j_3-P_SC^j\,(N^j_3-N^j)$ are constants solely determined by the helicity properties of the transition amplitudes.
Then, the integrands for the transverse and longitudinal components of the MSV of the Daughter can be given with spherical coordinates as
\bea\label{integrandPC1}
 {S}_{\Lambda x}^{PC}({\bf p}_*) &=&\frac{3}{2j(j+1)}\Big[2S_{H* x} \Big( A + B \cos^2 \varphi_* \sin^2 \theta_* \Big) + B \Big( S_{H* z} 
\cos \varphi_* \sin 2 \theta_* \nonumber\\
&&\qquad\qquad\ + S_{H* y} \sin 2 \varphi_* \sin^2 \theta_* \Big) \Big],\nonumber\\
{S}_{\Lambda y}^{PC}({\bf p}_*) &=&\frac{3}{2j(j+1)}\Big[2S_{H* y} \Big( A + B \sin^2 \varphi_* \sin^2 \theta_* \Big) + B \Big( S_{H* z} 
 \sin \varphi_* \sin 2 \theta_* \nonumber\\
 &&\qquad\qquad\ ++ S_{H* x} \sin 2 \varphi_* \sin^2 \theta_* \Big) \Big],\nonumber\\
{S}_{\Lambda z}^{PC}({\bf p}_*) &=&\frac{3}{2j(j+1)}\Big[2S_{H* z} \Big( A + B \cos^2 \theta_* \Big) + B \Big( S_{H* x} \cos \varphi_* \sin 2 \theta_*\nonumber\\
&&\qquad\qquad\ ++ S_{H* y} \sin \varphi_* \sin 2 \theta_* \Big) \Big].
\eea
Inserting \eqref{harmonics} into these integrands with the help of the trigonometric function relations \eqref{phimother} and using the even-odd properties listed in Table.\ref{evenodd}, only the following terms are non-vanishing when the integrations over the solid angle are taken into account (see Appendix.2):
\bea\label{integrandPC2}
 {S}_{\Lambda x}^{PC}({\bf p}_*) &=&[h_2F{\cal A}+Bg_0\sin^2 \theta_*\cos 2 \psi]\sin 2 \varphi_\Lambda+{B\over2}l_2^+{\cal F}_{2}^-\sin^2 \theta_*\sin 4 \varphi_\Lambda,\nonumber\\
{S}_{\Lambda y}^{PC}({\bf p}_*) &=& \left[g_0F+\frac{B}{2} l_2^-{\cal F}_{2}^+  \sin^2 \theta_*\right]+\Big[g_2 F {\cal A}-B g_0 \sin^2 \theta_* \cos 2 \psi \Big]\cos 2 \varphi_\Lambda\nonumber\\
&& -\frac{B}{2} l_2^+{\cal F}_{2}^- \sin^2 \theta_*\cos 4 \varphi_\Lambda,\nonumber\\
{S}_{\Lambda z}^{PC}({\bf p}_*) &=&\left[2f_2\!\left( A \!+\! B \cos^2 \theta_* \right)\! {\cal A}
\!+\! \frac{B}{2}l_1^+(\mathcal{C}\cos\psi\!-\!\mathcal{D}\sin\psi) \sin 2\theta_* \right]\sin 2\varphi_\Lambda,
\eea
where we define the auxiliary functions as:
$$l_n^\pm=h_n\pm g_n,\qquad F=2A + B \sin^2 \theta_*,\qquad{\cal F}_{n}^\pm={\cal A}\cos n\psi\pm{\cal B}\sin n\psi.$$
So, both the single-$\varphi_H$ harmonics in the transverse components of the MSV of the Mother contribute to the LLP ${S}_{\Lambda z}^{PC}({\bf p}_*) $, while its local feature $\sim\sin 2\varphi_H$ is well inherited by the Daughter with the polarization $\sim\sin 2\varphi_\Lambda$. The decays also give rise to higher mode of harmonics to the TLPs of the Daughter, that is, $\sin 4 \varphi_\Lambda$ and $\cos 4 \varphi_\Lambda$, even though the primary ones of the Mother are only to $2\varphi_H$ harmonics. As both $h_1$ and $g_1$ vanish for primary $\Lambda$ with $p_{z*}=0$, we arrive at a conclusion: only even-time harmonics of $\varphi_\Lambda$ are relevant to midrapidity $\Lambda$ polarizations, even after collecting the feed-downs from the strong and EM decays of the primary Mothers.

In weak decay, the previous polarization transfer pattern \eqref{SPC} remains important with the coefficients defined as $A=\gamma_w/4$ and $B=(1-\gamma_w)/4$ now. However, more terms are involved in weak decay, that is the $\alpha_w$ and $\beta_w$ dependent terms listed in Table.\ref{poltran}. The $\alpha_w$ term is irrelevant to the initial polarization of the Mother, and the contributions to the transverse and longitudinal components of the MSV of the Daughter can be given directly as
\bea\label{integranda}
{S}_{\Lambda x}^{\alpha_w}({\bf p}_*)&=&{\alpha_w\over2}\sin \theta_*\cos\psi\cos \varphi_\Lambda,\nonumber\\
{S}_{\Lambda y}^{\alpha_w}({\bf p}_*)&=&{\alpha_w\over2}\sin \theta_*\cos\psi\sin \varphi_\Lambda,\nonumber\\
{S}_{\Lambda z}^{\alpha_w}({\bf p}_*)&=&0.
\eea
The explicit forms for the corresponding contributions from $\beta_w$ term are
\bea
{S}_{\Lambda x}^{\beta_w}({\bf p}_*) &=&{\beta_w}\Big( S_{H* y} \cos \theta_* - S_{H* z} \sin \varphi_* \sin \theta_* \Big),\nonumber\\
{S}_{\Lambda y}^{\beta_w}({\bf p}_*) &=&{\beta_w}\Big( S_{H* z} \cos \varphi_* \sin \theta_* - S_{H* x} \cos \theta_* \Big),\nonumber\\
{S}_{\Lambda z}^{\beta_w}({\bf p}_*) &=&{\beta_w}\Big( S_{H* x} \sin \varphi_* \sin \theta_*- S_{H* y} \cos \varphi_* \sin \theta_* \Big).
\eea
Then, by following a similar procedure as that for the strong and EM decays, the terms giving rise to finite contributions are just
\bea\label{integrandb}
{S}_{\Lambda x}^{\beta_w}({\bf p}_*) 
&=&{\beta_w\over4}\Big[(-f_2{\cal F}_1^+\sin\theta_*+g_1\mathcal{C} \cos \theta_*)\cos \varphi_\Lambda+f_2{\cal F}_1^-\sin\theta_*\cos 3\varphi_\Lambda\Big],\nonumber\\
{S}_{\Lambda y}^{\beta_w}({\bf p}_*) 
&=&{\beta_w\over4}\Big[(f_2{\cal F}_1^+\sin\theta_*-h_1\mathcal{C} \cos \theta_*)\sin \varphi_\Lambda+f_2{\cal F}_1^-\sin\theta_*\sin 3\varphi_\Lambda\Big],\nonumber\\
{S}_{\Lambda z}^{\beta_w}({\bf p}_*)
&=&{\beta_w\over4}\sin \theta_*(l_2^-{\cal F}_1^+\cos\varphi_\Lambda
-l_2^+{\cal F}_1^-\cos3\varphi_\Lambda ).
\eea
Notice that $\alpha_w$ and $\beta_w$ terms only give rise to odd-time harmonics of $\varphi_\Lambda$, contrary to the even ones in strong and EM decays. 

Thus, by gathering \eqref{integrandPC2},\eqref{integranda} and \eqref{integrandb}, the most general integrands for the transverse and longitudinal components of the MSV of the Daughter fed-down from a single decay in HICs are
\bea\label{integrandg}
{S}_{\Lambda i}({\bf p}_*) = {S}_{\Lambda i}^{PC}({\bf p}_*)+{S}_{\Lambda i}^{\alpha_w}({\bf p}_*)+{S}_{\Lambda i}^{\beta_w}({\bf p}_*),\qquad i=x,y,z.
\eea
For the transverse components of the $\Lambda$ polarization, the $\beta_w$ contributions are much less important than the $\alpha_w$ ones because of the smallness of the polarization coefficients $h,g$ and $f$, thus they are suppressed in the following discussions. For the longitudinal component, the $\beta_w$ term from weak decay breaks the pure $\sin 2\varphi_\Lambda$ polarization structure of the Daughter inherited from the strong and EM decays in principle. However, for the specific case with $\Lambda$ polarization, the relevant decay parameters for $\Xi^0$ and $\Xi^-$ are~\cite{Tanabashi:2018oca}:
\bea
\alpha_w^{\Xi^0}=-0.347,\ \beta_w^{\Xi^0}=\tan (0.366\pm0.209)\gamma_w^{\Xi^0},\ \gamma_w^{\Xi^0}=0.85,\nonumber\\
\alpha_w^{\Xi^-}=-0.392,\ \beta_w^{\Xi^-}=\tan (0.037\pm0.014)\gamma_w^{\Xi^-},\ \gamma_w^{\Xi^-}=0.89.\nonumber
\eea
So $\beta_w^{\Xi^-}/\gamma_w^{\Xi^-}=0.037\pm0.014$ is very small and the breaking effect can be safely neglected for the weak decay of $\Xi^-$; but $\beta_w^{\Xi^0}/\gamma_w^{\Xi^0}=0.158-0.648$, the breaking effect might be large for that of $\Xi^0$. If we assume $|h_2|,g_2\lesssim|f_2|/2$ which is always true in HICs~\cite{Karpenko:2016jyx}, the magnitudes of the integrated coefficients in front of $\cos\varphi_\Lambda$ and $\cos3\varphi_\Lambda$ are at least one order smaller than that of $\sin 2\varphi_\Lambda$ for the largest ratio: $\beta_w^{\Xi^0}/\gamma_w^{\Xi^0}=0.648$. The reason can be well understood by comparing the prefactors in the integrands: Keep only the dominant ${\cal A}$ related term in ${\cal F}_1^\pm$, the ratios between the prefactors are roughly
$${\beta_w^{\Xi^0}\over\gamma_w^{\Xi^0}+(1-\gamma_w^{\Xi^0})\cos^2\theta_*}{h_2\mp g_2\over 2f_2}\sin \theta_*\cos\psi.$$
Then, they are double trigonometric function suppressed especially by $\cos\psi$ when carrying out the integrations, besides the initial suppression by $\beta_w^{\Xi^0}/\gamma_w^{\Xi^0}$. Similar comparisons can also be applied to the TLPs, thus the contributions from $\beta_w$ term will be neglected for $\Lambda$ polarization in the following.

At sufficiently high energy, because of the approximate longitudinal boost invariance, we expect 
all the functions $g,h$ and $f$ in \eqref{harmonics} to be very weakly dependent on the rapidity $Y_H$.
As a consequence, compared to the other rapidity-even functions, the rapidity-odd functions $h_1$ and $g_1$ can be safely neglected as they vanish at midrapidity $Y_H=0$. Finally, by inserting \eqref{integrandPC2} and \eqref{integranda} into \eqref{decayspin}, the total transverse and longitudinal components of the MSV of $\Lambda$ can be put in simple forms as
\bea
{\bf S}_{\Lambda x}(\p_\Lambda^T) &=&{1\over2}\left(H_1^{\rm Tot}({\p_\Lambda^T})\cos \varphi_\Lambda+H_2^{\rm Tot}({\p_\Lambda^T})\sin 2 \varphi_\Lambda+H_4^{\rm Tot}({\p_\Lambda^T})\sin 4 \varphi_\Lambda\right)\nonumber\\
&=&{1\over2}\sum_{M=\Lambda}^H\left(H_1^{M}({\p_\Lambda^T})\cos \varphi_\Lambda+H_2^{M}({\p_\Lambda^T})\sin 2 \varphi_\Lambda+H_4^{M}({\p_\Lambda^T})\sin 4 \varphi_\Lambda\right)\nonumber\\
&\equiv&{1\over2}\left[R_{\Lambda_p}  h_2\sin 2 \varphi_\Lambda\!+\!\sum_HR_H\left(h_1^{H}\cos \varphi_\Lambda\!+\!h_2^{H}\sin 2 \varphi_\Lambda\!+\!h_4^{H}\sin 4 \varphi_\Lambda\right)\right],\\\label{totalx}
&&\!\!\!\!\!\!\!\!\!\!\!h_1^{H}({\p_\Lambda^T})={{\alpha_w^{H}\over {\cal N}_H}\int \di \Omega_* \; n({\bf p}_H) \left\| \frac{\partial {\bf p}_H}{\partial{\bf p}_*} \right\|\, 
	\sin \theta_*\cos\psi},\nonumber\\
&&\!\!\!\!\!\!\!\!\!\!\!h_2^{H}({\p_\Lambda^T})={{2\over{\cal N}_H}\int \di \Omega_* \; n({\bf p}_H) \left\| \frac{\partial {\bf p}_H}{\partial{\bf p}_*} \right\|\, 
	\left[h_2({\p}_H^T)F^H{\cal A}+B^Hg_0({\p}_H^T)\sin^2 \theta_*\cos 2 \psi\right]},\nonumber\\
&&\!\!\!\!\!\!\!\!\!\!\!h_4^{H}({\p_\Lambda^T})={{B^H\over {\cal N}_H}\int \di \Omega_* \; n({\bf p}_H) \left\| \frac{\partial {\bf p}_H}{\partial{\bf p}_*} \right\|\, 
l_2^+({\p}_H^T){\cal F}_{2}^-\sin^2 \theta_*},\nonumber\\
{\bf S}_{\Lambda y}(\p_\Lambda^T) &=&{1\over2}\left(G_0^{\rm Tot}({\p_\Lambda^T})+G_1^{\rm Tot}({\p_\Lambda^T})\sin \varphi_\Lambda+G_2^{\rm Tot}({\p_\Lambda^T})\cos 2 \varphi_\Lambda+G_4^{\rm Tot}({\p_\Lambda^T})\cos 4 \varphi_\Lambda\right)\nonumber\\
&=&{1\over2}\sum_{M=\Lambda}^H\!\!\left(G_0^{M}({\p_\Lambda^T})\!+\!G_1^{M}({\p_\Lambda^T})\sin \varphi_\Lambda\!+\!G_2^{M}({\p_\Lambda^T})\cos 2 \varphi_\Lambda\!+\!G_4^{M}({\p_\Lambda^T})\cos 4 \varphi_\Lambda\right)\nonumber\\
&\equiv&{1\over2}\Big[R_{\Lambda_p}  (g_0+g_2\cos 2 \varphi_\Lambda)+\sum_HR_H\Big(g_0^{H}+h_1^{H}\sin \varphi_\Lambda+g_2^{H}\cos 2 \varphi_\Lambda\nonumber\\
&&\quad-h_4^{H}\cos 4 \varphi_\Lambda\Big)\Big],\label{totaly}\\
&&\!\!\!\!\!\!\!\!\!\!\!g_0^{H}({\p_\Lambda^T})={1\over{\cal N}_H}{\int \di \Omega_* \; n({\bf p}_H) \left\| \frac{\partial {\bf p}_H}{\partial{\bf p}_*} \right\|\, 
	\left[2g_0({\p}_H^T)F^H+B^H l_2^-({\p}_H^T){\cal F}_{2}^+  \sin^2 \theta_*\right]},\nonumber\\
&&\!\!\!\!\!\!\!\!\!\!\!g_2^{H}({\p_\Lambda^T})={2\over{\cal N}_H}{\int \di \Omega_* \; n({\bf p}_H) \left\| \frac{\partial {\bf p}_H}{\partial{\bf p}_*} \right\|\, 
	\left[g_2({\p}_H^T) F^H {\cal A}-B^H g_0({\p}_H^T) \sin^2 \theta_* \cos 2 \psi \right]},\nonumber\\
{\bf S}_{\Lambda z}(\p_\Lambda^T) &=&{1\over2}F_2^{\rm Tot}({\p_\Lambda^T})\sin 2 \varphi_\Lambda={1\over2}\left[\sum_{M=\Lambda}^HF_2^{M}({\p_\Lambda^T})\right]\sin 2 \varphi_\Lambda\nonumber\\
&\equiv&{1\over2}\left[R_{\Lambda_p}  f_2 ({\p_\Lambda^T})+\sum_HR_Hf_2^H ({\p}_\Lambda^T)\right] \sin 2\varphi_\Lambda,\label{totalz}\\
&&\!\!\!\!\!\!\!\!\!\!\!f_2^H({\p}_\Lambda^T)={4\over{\cal N}_H}{\int \di \Omega_* \; n({\bf p}_H) \left\| \frac{\partial {\bf p}_H}{\partial{\bf p}_*} \right\|\, 
	f_2 ({\p}_H^T)\left( A^H \!+\! B^H \cos^2 \theta_* \right) {\cal A}}\nonumber
\eea
with the normalization
$${\cal N}_H=\int \di \Omega_* n({\bf p}_H)  \left\| \frac{\partial {\bf p}_H}{\partial {\bf p}_*} \right\|.$$
Here, $h^H ({\p}_\Lambda^T),g^H ({\p}_\Lambda^T)$ and $f^H ({\p}_\Lambda^T)$ are the polarization transfer coefficients from the Mother and $R$'s are the $\Lambda$ number fractions from different contribution channels: $R_{\Lambda_p}$ primary and $R_{H}$ secondary. Due to the $2\pi$-periodicity of all the components with respect to $\varphi_\Lambda$, we can fold the transverse ones ${\bf S}_{\Lambda x}(\p_\Lambda^T)$ once over the region $\varphi_\Lambda\in(-\pi/2,3\pi/2)$ to $(-\pi/2,\pi/2)$ and ${\bf S}_{\Lambda y}(\p_\Lambda^T)$ twice over the region $\varphi_\Lambda\in(0,2\pi)$ to $(0,\pi/2)$, respectively. Then, all the trivial harmonics of $\varphi_\Lambda$ contributed from $\alpha_w^{H}$ terms will be removed from \eqref{totalx} and \eqref{totaly}, and the even-time harmonics of $\varphi_\Lambda$ can be explored in advance. For cascade decays, the evaluations of the MSV of the last Daughter should be done step by step, that is, iterating \eqref{totalx}, \eqref{totaly} and \eqref{totalz} over and over until the Daughter we're interested in. Take the EM decay $\Sigma^0\rightarrow\Lambda\gamma$ for example, we should first obtain the total polarization coefficients for $\Sigma^0$ including both the primary contributions and feed-downs from higher-lying resonances. Then, these total polarization coefficients, instead of the primary ones, are used to evaluate the contribution of $\Sigma^0$ decay to $\Lambda$ polarization, see \cite{Becattini:2016gvu,Xia:2019fjf} for numerical calculations.
\subsubsection{Theoretical predictions and sign puzzles}\label{puzzles}

In this section, we perform numerical calculations by adopting \eqref{totalx}, \eqref{totaly} and \eqref{totalz}, and compare the results with experimental measurements if available. In \cite{Becattini:2019ntv}, we just focused on the most important feed-down effects on the LLP of the $\Lambda$, that is, from the strong and EM decay channels with the Mother $H=\Sigma^*$ and $H=\Sigma^0$, respectively. A more complete study of all decay channels had been performed in \cite{Xia:2019fjf} and the conclusion remains the same for LLP. As mentioned before, these two parity conservative channels correspond to the decay types ${3/2^+}\rightarrow{1/2^+}0^-$ and ${1/2^+}\rightarrow{1/2^+}1^-$, and the decay coefficients are respectively
$$A^{\Sigma^*}=1/2,\ B^{\Sigma^*}=-1/4;\qquad A^{\Sigma^0}=0,\ B^{\Sigma^0}=-1/4.$$
The fractions of primary and secondary $\Lambda$ can be estimated by means of the statistical hadronization 
model. At the hadronization temperature $T=164$ MeV and baryon chemical potential of 30 MeV 
for $\snn =200$ GeVAu+Au collisions, they turn out to be~ \cite{Becattini:2016xct}: 
\bea
R_{\Lambda_p}= 0.243,\ R_{\Sigma^*} = 0.359,\ R_{\Sigma^0}=0.275*60\%,\label{fraction}
\eea
where $60\%$ is the contribution fraction from primary $\Sigma^0$ and the left from higher-lying resonance decays is assumed to cancel out for simplicity. At this hadronization temperature, the quantum statistics effects are negligible for all these particles, so the Boltzmann distinguishable particle assumption adopted in Sec.\ref{SDMM} is an excellent approximation. 

To perform numerical evaluations for the longitudinal component of the MSV of the Daughter \eqref{totalz}, two ingredients are still unknown: the primary LLP prefactor $f_2 ({\p^T})$ and the momentum spectrum $n({\bf p}_H)$. A precise fit to the data obtained in 
\cite{Becattini:2016gvu} for $f_2(\p_\Lambda^T)$ of the primary $\Lambda$ yields:
$$
f_2(\p_\Lambda^T)=\left[-7.71 \left(\p_\Lambda^T\right)^2+ 3.32  \left(\p_\Lambda^T\right)^3 - 0.471\left(\p_\Lambda^T\right)^4\right]\times 10^{-3}
$$
with $\p_\Lambda^T$'s unit "${\rm GeV}$". As far as $n({\bf p}_H)$ is concerned, it is plausible that the dependence on its form is very mild, because it shows in both the numerator and denominator of $f_2^H(\p_\Lambda^T)$. For the purpose of approximate calculations, we have assumed a spectrum of the following form~\cite{Becattini:2019ntv}:
\be\label{distribution}
n({\bf p}_H) \propto \frac{1}{\cosh Y_H} \e^{-m_H^T/T_s} = \frac{m_H^T}{\varepsilon_H} \e^{-m_H^T/T_s},
\ee
where $T_s$ is a phenomenological parameter describing the slope of the transverse momentum spectrum. It had been checked that the final results are almost independent of $T_s$ within a realistic range:
$T_s=0.2-0.8~{\rm GeV}$.

\begin{figure}[!htb]
	\sidecaption
	\includegraphics[scale=.65]{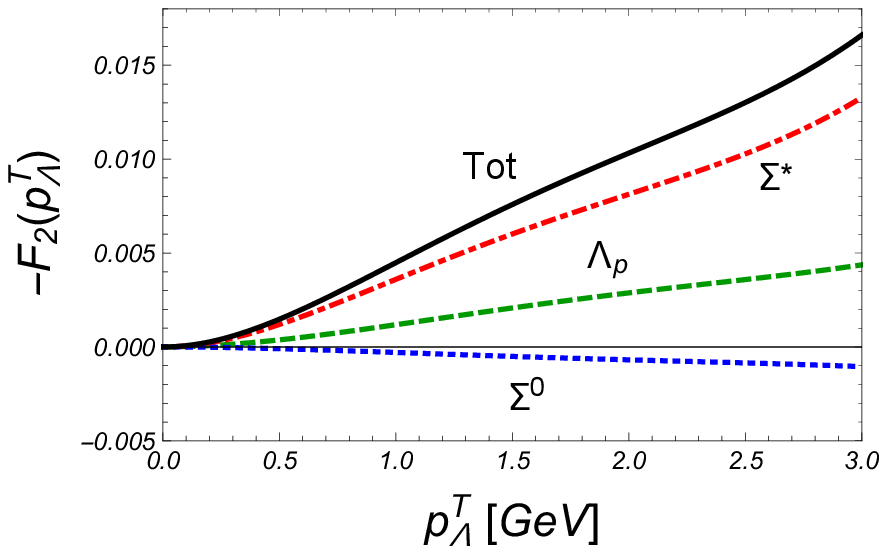}
	\includegraphics[scale=.65]{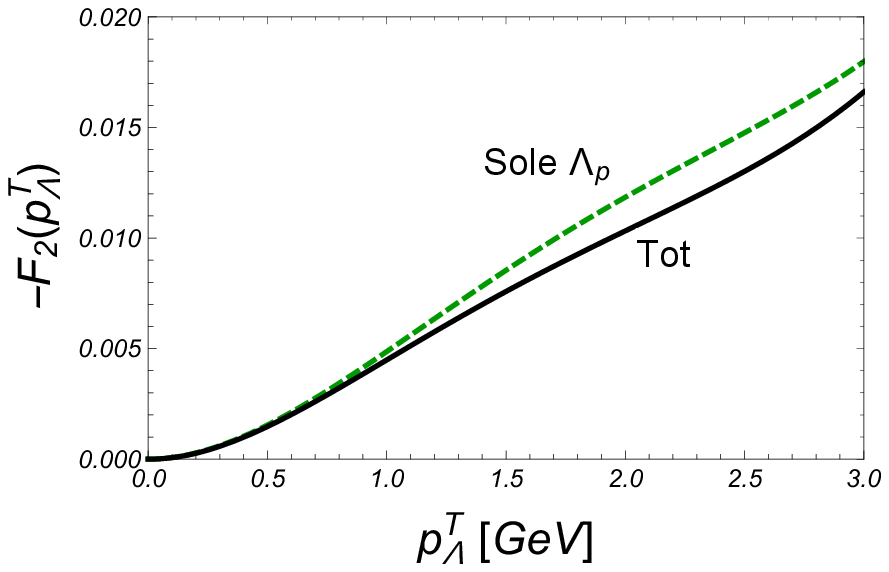}
	\caption{(color online) Left panel: longitudinal polarization coefficients $F_2(\p_\Lambda^T)$ of the $\Lambda$.
		Primary ($\Lambda_p$) and secondary ($H=\Sigma^*,\Sigma^0$) components, weighted with the production fractions are shown together with the resulting sum $F_2^{\rm Tot}(\p_\Lambda^T)$ (solid line). Right panel: comparison between the total polarization coefficient $F_2^{\rm Tot}(\p_\Lambda^T)$ of the $\Lambda$ and the one $f_2(\p_\Lambda^T)$ of only primary $\Lambda$~\cite{Becattini:2017gcx}. }
	\label{f2}     
\end{figure}
\begin{figure}[!htb]
	\sidecaption
	\includegraphics[scale=.65]{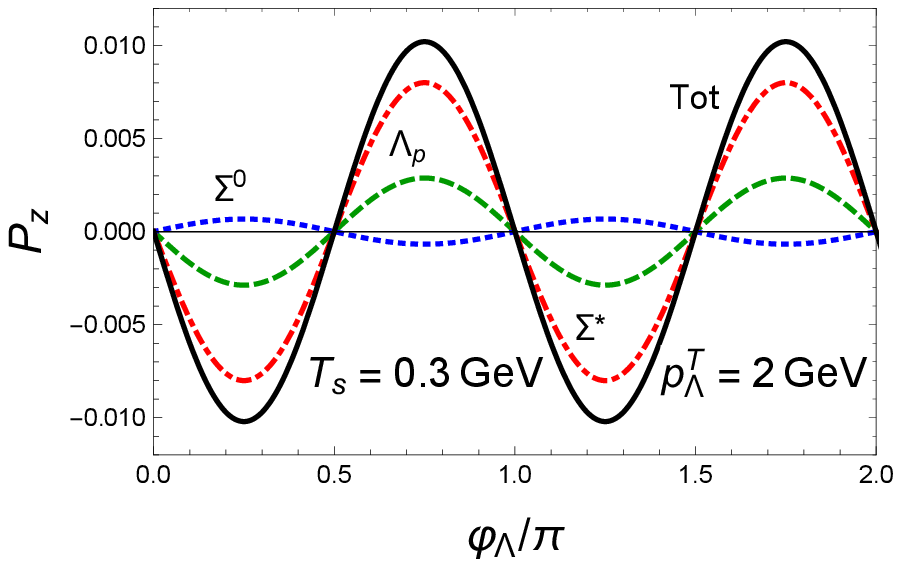}
	\includegraphics[scale=.65]{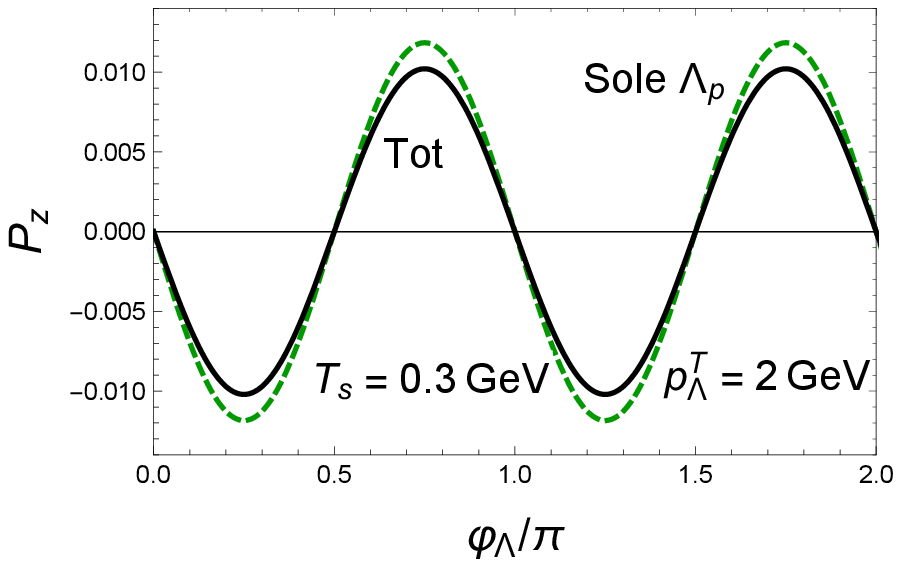}
	\caption{(color online) Left panel: the azimuthal angle dependence of the longitudinal polarization 
		$P_z ={\bf S}_{\Lambda z}(\p_\Lambda^T)/S = 2{\bf S}_{\Lambda z}(\p_\Lambda^T)$ of the $\Lambda$. Primary ($\Lambda_p$) and secondary ($H=\Sigma^*,\Sigma^0$) components, weighted with the production fractions are shown together with the resulting sum (solid line) at fixed transverse momentum $\p_\Lambda^T=2~{\rm GeV}$ and 
		slope parameter $T_s=0.3~{\rm GeV}$. Right panel: comparison between the total polarization profile of 
		the $\Lambda$ and that of only primary $\Lambda$ ~\cite{Becattini:2017gcx}.}
	\label{tot1}     
\end{figure}
\begin{figure}[!htb]
	\sidecaption
	\includegraphics[scale=.18]{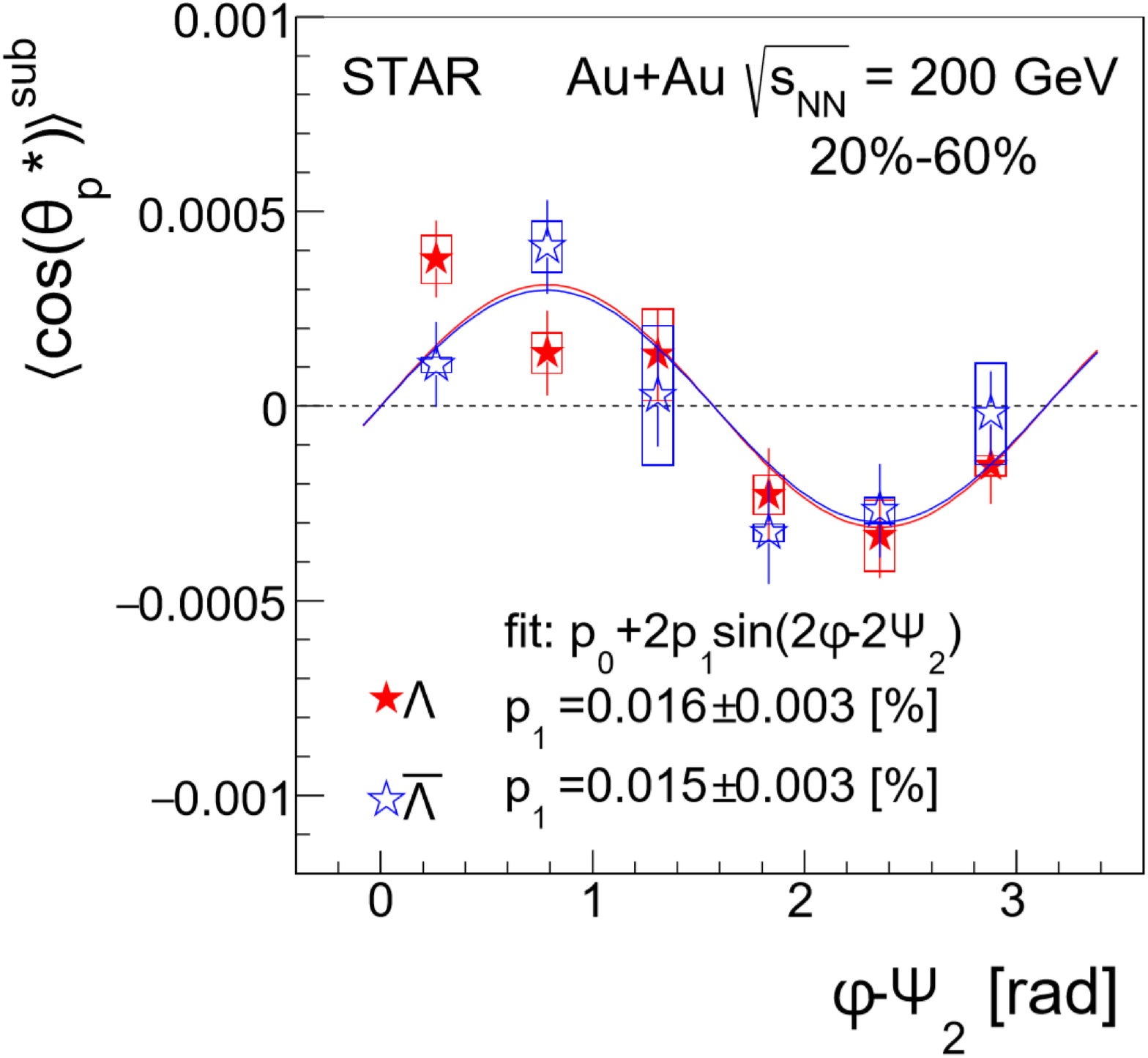}
	\caption{(color online) The experimental measurements of the longitudinal local polarizations of $\Lambda$ and $\bar\Lambda$ hyperons as functions of the azimuthal angle $\varphi$ relative to the second-order event plane $\Psi_2$ for $20\%-60\%$ centrality bin in $\sqrt{s_{NN}} = 200 {\rm GeV}$ Au+Au collisions~\cite{Niida:2018hfw,Adam:2019srw}. Solid lines show the fit with the function $\sin(2(\varphi-\Psi_2))$.}
	\label{expLLP}     
\end{figure}

The relevant polarization prefactors $F_2^M(\p_\Lambda^T)$ for primary and secondary decay components and the total $F_2^{\rm Tot}({\p_\Lambda^T})$ are shown together in Fig.~\ref{f2}, and the associated LLP features are illuminated in Fig.~\ref{tot1} where we choose $\p_\Lambda^T=2~{\rm GeV}$ as an example. As expected from the polarization transfer coefficients list in Table.\ref{poltran} and the fractions in \eqref{fraction}, the strong and EM decays give large positive and small negative feedbacks to the primary $\Lambda$ polarization, respectively, see the left panel in Fig.~\ref{f2}. It happens that $F_2^{\rm Tot}({\p_\Lambda^T})$ is close to the primary $f_2(\p_\Lambda^T)$ and only slightly suppressed in large $\p_\Lambda^T$ region, see the right panel in Fig.~\ref{f2}. In principle, there are also feedbacks from EM decay of secondary $\Sigma^0$ and weak decays of $\Xi$'s (positive), but their weights in $\Lambda$ productions are quite limited and definitely not able to flip the sign of $f_2(\p_\Lambda^T)$ in Fig.~\ref{f2}, see the results presented in \cite{Xia:2019fjf}. Compared to the theoretical predictions for the LLP profiles in Fig.~\ref{tot1}, the experimental measurements nicely verified the $\sin 2 \varphi_\Lambda$ feuture but with an opposite sign~\cite{Niida:2018hfw,Adam:2019srw}, see Fig.~\ref{expLLP} for both $\Lambda$ and $\bar{\Lambda}$ polarizations. We'd like to point out that this contradiction is not due to different conventions of the coordinate system in the theoretical and experimental studies. It is a real {\it sign puzzle} because the experimental measurements follow the same sign as that given by the differential of elliptic flow: $-\partial_\varphi v_2(\varphi)$~\cite{Niida:2018hfw,Adam:2019srw} but the theoretical predictions give opposite sign due to the negative prefactor $\di T/\di \tau$~\cite{Becattini:2017gcx}. Of course, this statement bases on the fact that the model calculations could well reproduce the elliptic flows measured in HICs, see hydrodynamic~\cite{Kolb:2000fha} and AMPT~\cite{Lin:2001zk} simulations for example.

\begin{figure}[!htb]
	\sidecaption
	\includegraphics[scale=.75]{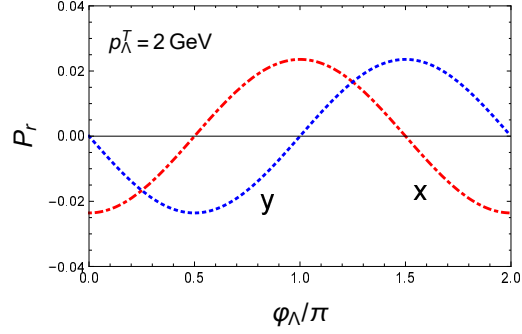}
	\caption{(color online) The radial polarizations $P_r^x$ (red dashed line) and $P_r^y$ (blue dotted line) of the $\Lambda$ as functions of the azimuthal angle at fixed transverse momentum $\p_\Lambda^T=2~{\rm GeV}$. Only the dominate $\alpha_w$ term is adopted for illumination.}
	\label{radial}     
\end{figure}

\begin{figure}[!htb]
	\sidecaption
	\includegraphics[scale=.75]{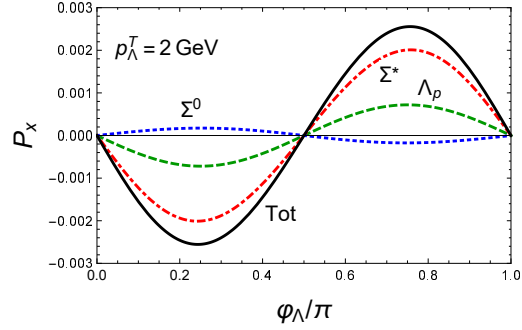}
	\caption{(color online) The azimuthal angle dependence of the folded transverse polarization 
		$P_x = 2{\bf S}_{\Lambda x}(\p_\Lambda^T)$ of the $\Lambda$. The parameters and denotations are the same as Fig.\ref{trany}.}
	\label{tranx}     
\end{figure}

For the TLPs, radial component $P_r$ was discovered mainly due to the parity-violating effect from weak decays~\cite{Xia:2019fjf}. According to \eqref{totalx} and \eqref{totaly}, the radial polarization of $\Lambda$ should be approximately proportional to $\alpha_w^{\Xi}R_{\Xi}$ with $R_{\Xi}\sim15\%$~\cite{Xia:2019fjf}. The results are shown in Fig.~\ref{radial} for $\p_\Lambda^T=2~{\rm GeV}$, where the $-\cos\varphi_\Lambda$ and $-\sin\varphi_\Lambda$ features are just inherited from the sign of $\alpha_w$, see \cite{Xia:2019fjf} for more realistic calculations. 
Now getting rid of the spontaneous radial polarization, we focus on the folded TLPs with the feed-down effect of the form \eqref{integrandPC2}. First of all, the folded results for ${\bf S}_{\Lambda x}^{PC}(\p_\Lambda^T)$ is studied and shown in Fig.~\ref{tranx}, where very nice $2\varphi_\Lambda$ harmonics can be identified. The higher harmonic $\sim\sin4\varphi_\Lambda$ vanishes here because the chosen parameters satisfy $h_2+g_2=0$ and we've checked that this contribution is very weak even for $g_2=h_2=-f_2/4$. 

\begin{figure}[!htb]
	\sidecaption
	\includegraphics[scale=.65]{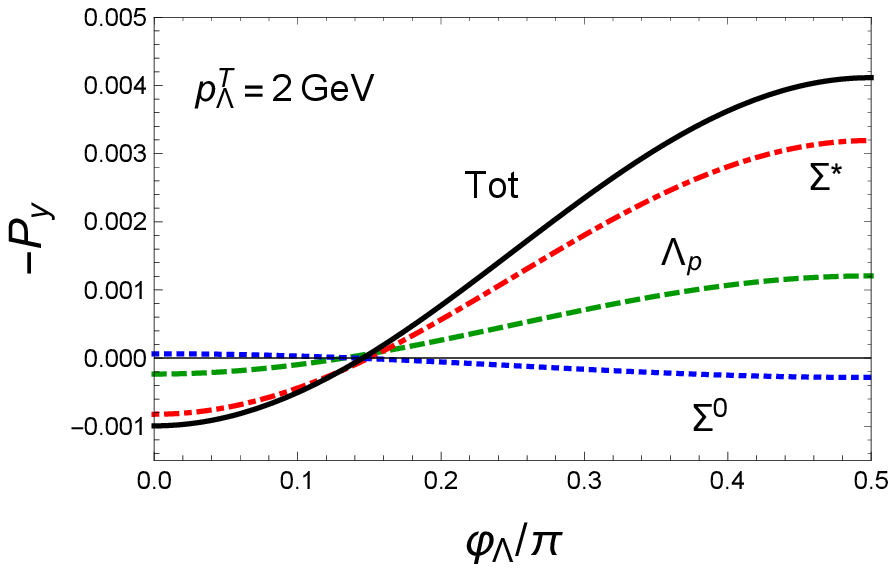}
	\includegraphics[scale=.35]{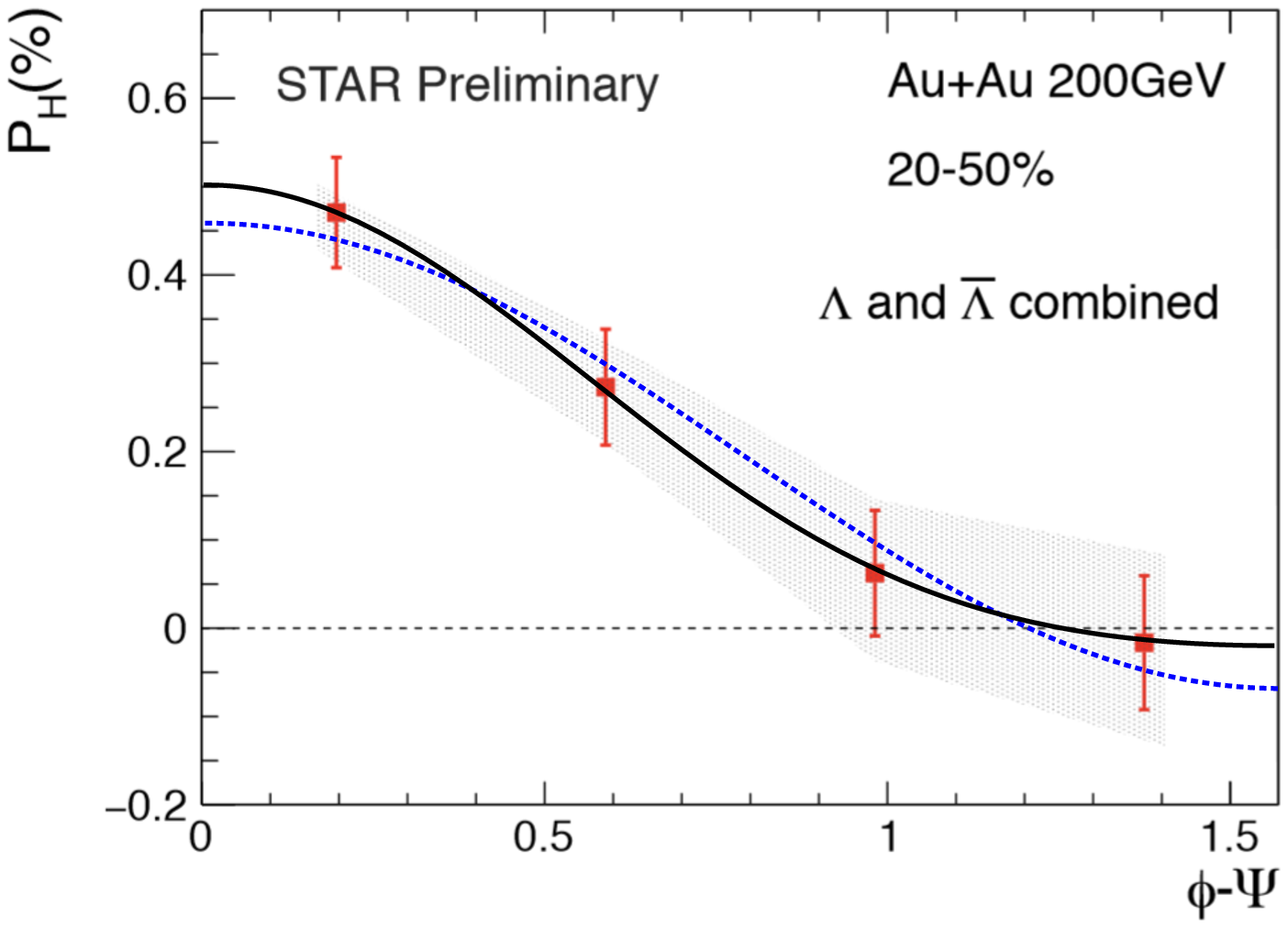}
	\caption{(color online) Left panel: the azimuthal angle dependence of the folded polarization along total angular momentum
		$P_y = 2{\bf S}_{\Lambda z}(\p_\Lambda^T)$ of the $\Lambda$. The parameters and denotations are the same as Fig.\ref{tot1}, and we choose $g_0=-0.004$ and $g_2=-h_2=-f_2/4$ according to the simulations in \cite{Karpenko:2016jyx}. Right panel: the experimental measurements of the polarizations of $\Lambda$ and $\bar\Lambda$ hyperons as functions of the azimuthal angle $\varphi$ relative to the first-order event plane $\Psi$ for $20\%-50\%$ centrality bin in $\sqrt{s_{NN}} = 200 {\rm GeV}$ Au+Au collisions~\cite{Niida:2018hfw}. Solid and dotted lines show the fits with even cosine harmonics up to quadruple and double angles, respectively.}
	\label{trany}     
\end{figure}

For the more involved TLP ${\bf S}_{\Lambda y}^{PC}(\p_\Lambda^T)$, the comparison between theoretical predictions for $\p_\Lambda^T=2~{\rm GeV}$ and experimental measurements is illuminated in Fig.~\ref{trany}. Due to different conventions for $y$-axis, the polarization $-P_y$ predicted in theoretical study corresponds to $P_H$ measured in experiments. Then, we immediately find that the signs of the azimuthal angle averaged $-P_y$ and the global $P_H$ are consistent with each other, which just follow that of the total angular momentum. However, the relative magnitudes between the in-plane ($\varphi_\Lambda=0$) and out-plane ($\varphi_\Lambda=\pi/2$) polarizations are opposite in the theoretical and experimental studies. The theoretical profile originates from the opposite signs between $g_0$ and $g_2$ as discussed in \eqref{signs} and the feed-down effect from the Mothers would not change that, see also \cite{Xia:2019fjf}. So, this is another {\it sign puzzle} in $\Lambda$ polarization and definitely rules out the naiive guess that the contradictions between theoretical and experimental results are only due to different conventions of the coordinate system. 

As indicated in \eqref{integrandPC2}, secondary decays can give rise to $4\varphi_\Lambda$ harmonic of $\Lambda$ polarization along the total angular momentum even though only up-to $2\varphi_H$ harmonics of the primary Mother polarizations are considered.  Similar to ${\bf S}_{\Lambda x}^{PC}(\p_\Lambda^T)$,  this higher harmonic vanishes in the left panel of Fig.~\ref{trany} because of the choice $h_2+g_2=0$ and this contribution is still very weak even for $g_2=h_2=-f_2/4$. We give the best fits to the experimental data in the right panel of Fig.~\ref{trany}: though the fit with up to $4\varphi_\Lambda$ harmonics has more advantage to reproduce the central values, the fit with up to $2\varphi_\Lambda$ harmonics is also consistent with the data within errorbars. 

We conclude that while the theoretical predictions and the experimental measurements are consistent with each other for the global polarization of $\Lambda$, the azimuthal angle dependences for either the longitudinal and transverse polarizations give opposite signs. Though the component ${\bf S}_{\Lambda x}^{PC}(\p_\Lambda^T)$ has not been measured in experiments, we expect the sign to be also opposite to the theoretical one, which then shares the same origin as the previous {\it sign puzzles}. Taking into account the feed-down effect of higher-lying hyperon decays~\cite{Becattini:2019ntv,Xia:2019fjf}, the final amplitudes of the $2\varphi_\Lambda$ harmonics are almost the same as that given by the primary $\Lambda$. Thus, sign flips are still impossible even after taking into account the contributions from resonance decays. Compared to the global polarization, the local polarizations always involve the thermal vorticity with time component (TVWTC)~\cite{Karpenko:2016jyx,Becattini:2017gcx}, so the answers to the {\it sign puzzles} might be closely related to this component. Actually, several definitions of vorticity~\cite{Becattini:2015ska} including "thermal", "kinematic" and "temperature" ones are compared in \cite{Wu:2019eyi}: The kinematic one gives the same signs as the thermal one which indicates the overwhelming role of VWTC, while the temperature one gives the correct signs as the experiments because its dependence on temperature is inverse to that of thermal one. Besides, getting rid of the TVWTC, the sign was found to be consistent with experimental measurements for the LLP of $\Lambda$~\cite{Adam:2019srw,Florkowski:2019voj}. In this models, the opposite effects seems to be simply originated from the opposite contributions of $\varpi_{01} ,\varpi_{02}~(<0)$ and $\varpi_{12}~(>0)$ to the MSV of the hyperons in the last equality of \eqref{vortspin}. However, even the hydrodynamic simulations, following the non-relativistic definition of vorticity, doesn't give the same sign as the experimental measurement. Thus, the reason is  not so trivial. We have a better proposal: it might be the higher-order derivative corrections to the commonly adopted thermal vorticity that change the whole features.

\section*{Appendix 1 $\ $ Lorentz boost and Jacobian determinant}
\label{app:lorentz}
\addcontentsline{toc}{section}{Appendix 1 $\ $ Lorentz transformation and Jacobian determinant}

In this Appendix, we demonstrate details to derive \eqref{momrel} and \eqref{jacob} shown in Sec. \ref{Momentumaverage}. 
As mentioned in the context, $p^\mu_\Lambda=(\varepsilon_\Lambda, {\bf p}_\Lambda)$ and $p_*^\mu=(\varepsilon_{\Lambda*} , {\bf p}_*)$ are the four-momenta of $\Lambda$ in QGP frame and Mother's rest frame, respectively, and $p_H^\mu=(\varepsilon_H, {\bf p}_H)$ 
the four-momentum of the Mother in QGPF. The pure Lorentz boost transforming the 
momentum of $\Lambda$ from QGPF to MRF reads:
\begin{align}
\varepsilon_{\Lambda*}={}&\gamma_H(\varepsilon_\Lambda -{\bf v}_H\cdot {\bf p}_\Lambda),\label{lorr1} \\
{\bf p}_* = {}& {\bf p}_\Lambda + \left(\frac{\gamma_H -1}{{\bf v}_H^2}{\bf v}_H \cdot {\bf p}_\Lambda - 
\gamma_H~\varepsilon_\Lambda \right){\bf v}_H ,
\label{lorr2}
\end{align}
where ${\bf v}_H={\bf p}_H/\varepsilon_H$ is the velocity of the Mother and $\gamma_H=\varepsilon_H/m_H$ 
the corresponding Lorentz factor. Hence, the explicit forms of \eqref{lorr1} and \eqref{lorr2} are:
\begin{align}
\varepsilon_{\Lambda*}={}&{1\over m_H} (\varepsilon_H\varepsilon_\Lambda - {\bf p}_H\cdot {\bf p}_\Lambda),\label{lorr11} \\
{\bf p}_* = {}& {\bf p}_\Lambda + \left[ \frac{{\bf p}_H\cdot{\bf p}_\Lambda}{m_H (\varepsilon_H + m_H)} -\frac{\varepsilon_\Lambda}{m_H} \right]{\bf p}_H,
\label{lorr22}
\end{align}
then the expression of ${\bf p}_H\cdot {\bf p}_\Lambda$ from \eqref{lorr11} can be substituted into \eqref{lorr22} to get
\be\label{ppp}
{\bf p}_* = {\bf p}_\Lambda + \left[ \frac{\varepsilon_H\varepsilon_\Lambda - m_H \varepsilon_{\Lambda*}}{m_H(\varepsilon_H+m_H)}  -\frac{\varepsilon_\Lambda}{m_H} \right]{\bf p}_H
= {\bf p}_\Lambda - \frac{\varepsilon_{\Lambda*} + \varepsilon_\Lambda}{\varepsilon_H+m_H}{\bf p}_H.
\ee
Move ${\bf p}_\Lambda$ to the left-hand side of \eqref{ppp} and take square of both sides, we have
\be\label{pmp}
({\bf p}_* - {\bf p}_\Lambda)^2=\frac{(\varepsilon_{\Lambda*} + \varepsilon_\Lambda)^2}{(\varepsilon_H+m_H)^2}{\bf p}_H^2 
= \frac{\varepsilon_H-m_H}{\varepsilon_H+m_H}(\varepsilon_{\Lambda*} + \varepsilon_\Lambda)^2,
\ee
which then gives the energy of the Mother in terms of the energy-momenta of the Daughter as:
\be
\label{EEE}
\varepsilon_H=m_H \frac{(\varepsilon_{\Lambda*} + \varepsilon_\Lambda)^2 + ({\bf p}_* - {\bf p}_\Lambda)^2}{(\varepsilon_{\Lambda*} + \varepsilon_\Lambda)^2 
	- ({\bf p}_* - {\bf p}_\Lambda)^2}.
\ee
By substituting \eqref{EEE} back into \eqref{ppp}, the final expression for the momentum of the Mother follows directly:
\be\label{pH}
{\bf p}_H= 2m_H \frac{\varepsilon_+{\bf p}_-}{\varepsilon_+^2- {\bf p}_-^2} \quad\text{with}\quad \varepsilon_+=\varepsilon_\Lambda+\varepsilon_{\Lambda*}, \quad
{\bf p}_-={\bf p}_\Lambda-{\bf p}_*.
\ee

Now, the above equation \eqref{pH} can be easily adopted to alter the integration variable involved in \eqref{Mvar} from ${\bf p}_H$ to ${\bf p}_*$ by fixing ${\bf p}_\Lambda$. The Jacobian matrix of the transformation can be evaluated as: 
\be
\frac{\partial {\rm p}_{Hi}}{\partial \p_{*j}}=\frac{2m_H}{\varepsilon_+^2-{\bf p}_-^2} \left\{\left[\p_{- ,i}\frac{\p_{*j}}{\varepsilon_{\Lambda*}} 
- \varepsilon_+\delta_{ij} \right]-{2\varepsilon_+\p_{- ,i}\over\varepsilon_+^2-{\bf p}_-^2} \left[\varepsilon_+\frac{\p_{*j}}{\varepsilon_{\Lambda*}} 
+\p_{- ,j} \right]\right\}
\ee
for $i,j=x,y,z$, and the determinant follows directly after some algebraic manipulations:
\be
\left| \frac{\partial {\bf p}_H}{\partial{\bf p}_*} \right|= \frac{4m_H^3\varepsilon_+^2(\varepsilon_+^2+{\bf p}_-^2)}{\varepsilon_{\Lambda*}(\varepsilon_+^2-{\bf p}_-^2)^3}.
\ee
%

\section*{Appendix 2 $\ $ Integrands for the transverse and longitudinal polarizations}\label{App:Integrands}
\addcontentsline{toc}{section}{Appendix 2 $\ $ Integrands for the transverse and longitudinal polarizations}

Herein, we work out the integrands for the evaluations of the transverse and longitudinal components of the mean spin vector, fed down from the strong and EM decays. Take the most complicated component ${S}_{\Lambda y}^{PC}({\bf p}_*)$, along the total angular momentum, for example, inserting \eqref{harmonics} into the second equation of \eqref{integrandPC1} gives
\bea\label{trapol0}
{S}_{\Lambda y}^{PC}({\bf p}_*)&=&2(g_0 + g_1 \cos \varphi_H 
+ g_2 \cos 2 \varphi_H) \Big( A + B \sin^2 \varphi_* \sin^2 \theta_* \Big)  + B \Big( f_2\sin 2 \varphi_H\nonumber\\
&&
\sin \varphi_* \sin 2 \theta_* + (h_1 \sin \varphi_H
+ h_2 \sin 2 \varphi_H) \sin 2 \varphi_* \sin^2 \theta_* \Big).
\eea
Because $h_1({\rm P}_T,Y_H)$ and $g_1({\rm P}_T,Y_H)$ are odd functions of $Y_H$ thus also of "$\cos\theta_*$" and all the trigonometric functions of the Mother in \eqref{phimother} are even functions of "$\cos\theta_*$", the terms proportional to $h_1$ and  $g_1$ do not contribute at all after integrating over $\theta_*$. Likewise, the term proportional to $f_2({\rm P}_T,Y_H)$, 
which is an even function of "$\cos\theta_*$", vanishes upon integration over $\theta_*$ 
because the function $\sin 2 \theta_*$ is odd. So we are left with:
\be\label{trapol1}
{S}_{\Lambda y}^{PC}({\bf p}_*)=(g_0 + g_2 \cos 2 \varphi_H) \left(F - B\cos 2\varphi_*\sin^2 \theta_* \right) + B\, h_2 \sin 2 \varphi_H \sin 2 \varphi_* \sin^2 \theta_*, 
\ee
where $F=2A+B\sin^2 \theta_*$.

Insert \eqref{phimother} and replace $\varphi_*$ by $ \varphi_\Lambda + \psi$, \eqref{trapol1} becomes explicitly
\bea\label{trapol2}
&&[g_0\! +\! g_2 ({\cal A}\cos 2\varphi_\Lambda\! -\! {\cal B}\sin2\varphi_\Lambda)] \left[F\! -\! B(\cos 2\varphi_\Lambda\cos2\psi\!-\!\sin 2\varphi_\Lambda\sin2\psi)\sin^2 \theta_* \right]\nonumber\\
&& + B\, h_2({\cal A}\sin2\varphi_\Lambda+{\cal B}\cos 2\varphi_\Lambda) (\cos 2\varphi_\Lambda\sin2\psi+\sin 2\varphi_\Lambda\cos2\psi) \sin^2 \theta_*.
\eea
Remember that any terms that are odd functions of "$\cos\theta_*$" or $\psi$ vanish after solid angle integrations. Thus, by taking into account the even-oddness of the relevant functions listed in Table.\ref{evenodd}, the following terms are left:
\bea\label{trapol3}
&&(g_0\! +\! g_2 {\cal A}\cos 2\varphi_\Lambda)(F\! -\! B\cos 2\varphi_\Lambda\cos2\psi \sin^2 \theta_*)\! -\!g_2 {\cal B}B\sin^22\varphi_\Lambda\sin2\psi\sin^2 \theta_*\nonumber\\
&& + {B} h_2({\cal A}\sin^22\varphi_\Lambda\cos2\psi+{\cal B}\cos^2 2\varphi_\Lambda\sin2\psi)\sin^2 \theta_*.
\eea
Finally, we adopt the double-angle relationships for the trigonometric functions:
$$\cos^2x={1\over2}(\cos2x+1),\qquad\sin^2x={1\over2}(-\cos2x+1)$$
to put the result \eqref{trapol3} in harmonics of $\varphi_\Lambda$:
\bea\label{trapol4}
{S}_{\Lambda y}^{PC}({\bf p}_*)&=&\left[g_0F+{B\over2} (h_2-g_2) ({\cal A}\cos2\psi+{\cal B}\sin2\psi) \sin^2 \theta_*\right]\! \nonumber\\
&&-(g_0B\cos 2\psi \sin^2 \theta_*-g_2F {\cal A})\cos2\varphi_\Lambda\nonumber\\
&& - {B\over2} (h_2+g_2)({\cal A}\cos2\psi-{\cal B}\sin2\psi)\sin^2 \theta_*\cos4\varphi_\Lambda.
\eea
One finds that $h_2$ and $g_2$ terms give rise contributions to both global and $4\varphi_\Lambda$ harmonic modes for the TLP $P_y$.

Similarly, $h_1,g_1$ and $f_2$ do not contribute to the TLP $P_x$ because the relevant terms in the integrand ${S}_{\Lambda x}^{PC}({\bf p}_*)$ are also odd functions of "$\cos\theta_*$". So by combining \eqref{harmonics} and \eqref{phimother} with the first equation in \eqref{integrandPC1}, the integrand is explicitly	
\bea
{S}_{\Lambda x}^{PC}({\bf p}_*)&=& h_2({\cal A} \sin 2 \varphi_\Lambda + {\cal B} \cos 2 \varphi_\Lambda)\Big( F + {B} \cos2 \varphi_* \sin^2 \theta_* \Big)\nonumber\\
&& + B [g_0+g_2({\cal A} \cos 2 \varphi_\Lambda - {\cal B} \sin 2 \varphi_\Lambda)] \sin 2 \varphi_* \sin^2 \theta_*,
\eea
which becomes
\bea
&& h_2\left[{\cal A} \sin 2 \varphi_\Lambda \Big( F + {B} \cos2 \psi\cos 2 \varphi_\Lambda \sin^2 \theta_* \Big)-{B\over2} {\cal B}\sin2 \psi \sin 4 \varphi_\Lambda\sin^2 \theta_* \right]\nonumber\\
&& + B \left[(g_0+g_2{\cal A} \cos 2 \varphi_\Lambda)\cos2 \psi \sin 2 \varphi_\Lambda -g_2 {{\cal B}\over2}\sin4\varphi_\Lambda \sin2 \psi\right] \sin^2 \theta_*.
\eea
after replacing $\varphi_*$ by $ \varphi_\Lambda + \psi$. And the double-angle relationships give
\bea
{S}_{\Lambda x}^{PC}({\bf p}_*)
&=&(h_2F{\cal A}+g_0B\cos 2 \psi\sin^2 \theta_*)\sin 2 \varphi_\Lambda\nonumber\\
&&+{B\over2}(h_2+g_2)({\cal A}\cos 2 \psi-{\cal B}\sin 2 \psi)\sin^2 \theta_*\sin 4 \varphi_\Lambda,
\eea
where we recognize that the coefficient of the $4\varphi_\Lambda$ harmonic is opposite to that of ${S}_{\Lambda y}^{PC}({\bf p}_*)$.

For the longitudinal component, $g_0,g_2$ and $h_2$ do not contribute because the relevant terms in the integrand ${S}_{\Lambda z}^{PC}({\bf p}_*)$ are also odd functions of "$\cos\theta_*$". So by combining \eqref{harmonics} and \eqref{phimother} with the third equation in \eqref{integrandPC1}, the integrand is explicitly
\bea
{S}_{\Lambda z}^{PC}({\bf p}_*)&=&2 f_2({\cal A} \sin 2 \varphi_\Lambda \!+\! {\cal B} \cos 2 \varphi_\Lambda)\Big(A\! +\! {B} \cos^2 \theta_* \Big)\!+\! B [h_1({\cal C} \sin \varphi_\Lambda \!+\! {\cal D} \cos \varphi_\Lambda)\nonumber\\
&& \cos \varphi_*+g_1({\cal C} \cos \varphi_\Lambda - {\cal D} \sin \varphi_\Lambda)\sin \varphi_*]  \sin 2 \theta_*,
\eea
which becomes
\be
{S}_{\Lambda z}^{PC}({\bf p}_*)=\left[2 f_2{\cal A}\Big(A\! +\! {B} \cos^2 \theta_* \Big)\!+\! {B\over2} (h_1+g_1)({\cal C}\cos \psi \!-\! {\cal D} \sin \psi) \sin 2 \theta_*\right]\sin 2 \varphi_\Lambda
\ee
after replacing $\varphi_*$ by $ \varphi_\Lambda + \psi$. Note that the LLP keeps the same harmonic as the primary one without any other mixing, that is, $\sim\sin 2 \varphi_\Lambda$.

\end{document}